\documentclass[journal]{IEEEtran}

\usepackage{xcolor}
\usepackage{graphicx}
\usepackage{subfigure}
\usepackage{soul}
\usepackage{amsmath}
\usepackage{amssymb}
\usepackage{amsthm}
\usepackage{amsbsy}
\usepackage{mathtools}
\usepackage{float}
\usepackage{cite}
\usepackage{tikz}

\usepackage{pbox}

\soulregister{\cite}7
\soulregister{\eqref}7
\soulregister{\ref}7

\DeclareMathOperator{\diag}{diag}

\newcommand\calF{{\mathcal{F}}}

\newcommand{\sgn}{\mathrm{sgn}}

\ifCLASSINFOpdf
\else
\fi
\begin{document}
%
\title{Implicit Euler Discrete-Time Set-Valued Admittance Control for Impact-Contact Force Control }
%
%
%

\author{
        Ke Li,
Xiaogang Xiong,~\IEEEmembership{Member,~IEEE,}
        Anjia Wang,
Ying Qu,
        and~Yunjiang Lou,~\IEEEmembership{Senior Member,~IEEE}
\thanks{
*This work was supported in part by the Shenzhen Science and Technology Program under KQTD20190929172545139 and GXWD20231130153844002, supported by the Open Fund of Laboratory of Aerospace Servo Actuation and Transmission, and partially supported by GuangDong Basic and Applied Basic Research Foundation No. 2022A1515011521;  (Corresponding authors: Xiaogang Xiong and Yunjiang Lou)}
\thanks{K. Li, X. Xiong, Y. Qu, A. Wang, and Y. Lou are with the School of Mechanical Engineering and Automation of Harbin Institute of Technology Shenzhen, Shenzhen 518055, China. (E-mail: \{keli, 22S153117,\, 21s053091\}@stu.hit.edu.cn, \{xiongxg,\,lourj\}@hit.edu.cn;) }
}

\maketitle

\begin{abstract}

Admittance control is a commonly used strategy for regulating robotic systems, such as quadruped and humanoid robots, allowing them to respond compliantly to contact forces during interactions with their environments. However, it can lead to instability and unsafe behaviors like snapping back and overshooting due to torque saturation from impacts with unknown stiffness environments. This paper introduces a novel admittance controller that ensures stable force control after impacting unknown stiffness environments by leveraging the differentiability of impact-contact forces. The controller is mathematically represented by a differential algebraic inclusion (DAI) comprising two interdependent set-valued loops. 
The first loop employs set-valued first-order sliding mode control (SMC) to limit input torque post-impact. The second loop utilizes the multivariable super-twisting algorithm (MSTA) to mitigate unstable motion caused by impact forces when interacting with unknown stiffness environments.
Implementing this proposed admittance control in digital settings presents challenges due to the interconnected structure of the two set-valued loops, unlike implicit Euler discretization methods for set-valued SMCs. To facilitate implementation, this paper offers a new algorithm for implicit Euler discretization of the DAI. Simulation and experimental results demonstrate that the proposed admittance controller outperforms state-of-the-art methods.

\end{abstract}


%
\IEEEpeerreviewmaketitle

\section{Introduction}
%
%
%
%


\IEEEPARstart{I}{mpact}-contact occurs in various scenarios where robots interact with their environments. For instance, the motion stability of quadruped robots relies on the force control of impact contact between their feet and the ground to maintain balance and support their bodies \cite{Xiong_2023_Quadruped}. Similarly, unmanned aerial manipulators (UAMs) must absorb and compensate for external impacts on their end-effectors to ensure compliant motion and maintain a fixed position when catching moving objects or payloads \cite{Haoyao_2024_RAL}. In these situations, robots need to adjust their movements after impact contact with unknown or changing environments, particularly regarding stiffness, even when actuation torque is saturated post-impact.

The admittance control can be regarded as one of general impedance control that regulates the relationship between the displacement or velocity of end-effector of robotic systems and external contact force during interactions with environments \cite{Calanca_2016_review,Kikuuwe_TRO_2019}. The typical impedance control regulates the motion of a proxy system with the designed mechanical characteristics (inertia, viscosity, and stiffness).
Comparing to the typical impedance control, the admittance control is also called ``position-based impedance control", because it internally contains a high-gain position controller. This high-gain controller suppresses the dynamics of robotic systems such as the nonlinear friction to achieve highly accurate target position tracking, and thus it has widely applications in compliance control, such as the manual teaching of industrial manipulators, interaction between human and robots, and surgical robots \cite{Zhijun_2022_TRO,Kang_2019_Tmech,Kim_2022_Mech}.

%

\par


The compliance motion of admittance control for robot-environment interactions has been widely concerned in different situations, as discussed in
the literature \cite{Kim_2022_Mech,Yinjie_2021_Tmech,Yinjie_2022_RAL}. The position of its internal proxy system in admittance control
can vibrate or diverge from the actual position of the controlled robot, caused by actuation saturation, unmeasurable force or contact with stiff environments. Due to its internal high-gain position controller, the vibrated or diverged proxy position can lead the admittance controller to produce unpredictable behaviors such as overshooting, snapping-back, and oscillations. These behaviors pose threats and safety problems to surrounding people and environments and the situation becomes worsen when the admittance control deals with impact-contact scenarios.
For example, as reported in \cite{Kikuuwe_2014_Tmech,Kikuuwe_TRO_2019}, during the motion of a robot manipulator, when it was impacted by a large force exerted away from its force sensor, undesirable behaviors can happen when the torque of the internal position control is saturated.
Another issues of admittance control is the instability when contacting with surrounding environments of high stiffness, which has been a long-standing problem for admittance control \cite{Tahara_2021_ICRA}.

\par

To solve these problems of the admittance control, different approaches were proposed in the literature \cite{Kikuuwe_2014_Tmech,Kikuuwe_TRO_2019,Tahara_2021_ICRA,Ueki_2021_ICM,Xiong_Tmech_2023}. Kikuwe \cite{Kikuuwe_2014_Tmech} proposed a sliding-mode-like internal position controller in the conventional admittance control to attenuate the undesirable behaviors caused by torque saturation, but it is only effective for short-time saturation, as pointed out in \cite{Kikuuwe_TRO_2019}. In \cite{Kikuuwe_TRO_2019,Xiong_Tmech_2023}, a nonsmooth set-valued normal cone operation was included into the proxy system of the conventional admittance controller. This method can suppress the unsafe behaviors during torque saturation, but they cannot deal with impact-contact scenarios. To attenuate the magnitudes of oscillations during contact with stiff objects, T. Fujiki and K. Tahara \cite{Tahara_2021_ICRA} serially combined the conventional impedance control and admittance control. However, this combined structure cannot remove the oscillation behaviors, and it still suffers from the safety problem when the actuator is saturated.

In \cite{Ueki_2021_ICM}, to avoid unstable robot behaviors of the impedance and admittance controls, a method of trajectory planning was proposed to restrain the contact force in contact-rich tasks with robots. In \cite{WeiHe_2020_TIE}, based on neural network approximations, an admittance control was proposed to compensate for unsafe behaviors when the robots' torque is saturated. In \cite{Yinjie_2021_Tmech} and \cite{Yinjie_2022_RAL}, a unified virtual proxy system was designed to control robots' motion, force, and impedance simultaneously and the parameters of proxy system changed according to different contact scenarios. Similarly to the approach in \cite{Chengguang_2022_TNNLS}, these learning or data-driven methods require sufficient interactions between robots and environments to collect data and learn about the environments before deployment. 
Unlike these learning-based approaches, T. Fujiki and K. Tahara \cite{Tahara_2021_ICRA} combine admittance control and impedance control in series in one feedback loop to improve environmental interaction adaptability. However, like impedance control, this method reduces oscillation at the expense of control accuracy.
The work in \cite{Kikuuwe_TRO_2019} employed a set-valued loop for the proxy system of an admittance controller, resulting in a safer control mechanism. Similarly, the authors' previous work \cite{Xiong_Tmech_2023} not only introduced a set-valued loop to the proxy system, but also to the inner position controller of an admittance control system, enhancing its safety and robustness to the system during interaction to environments. These studies employed implicit Euler discrete-time realizations of set-valued admittance control to make implementation feasible. However, the works \cite{Kikuuwe_TRO_2019,Xiong_Tmech_2023} are limited to systems with one degree of freedom (DoF) or multi-DoF systems that can be decoupled.

\par


This paper proposes a novel admittance control with two set-valued loops for coupled multi-DoF systems to ensure safe behavior during actuation saturation and contact with environments of unknown stiffness. The outer loop uses a proxy system with set-valued first-order sliding mode control (SMC) to suppress unsafe behaviors caused by actuation saturation due to environmental interaction. Meanwhile, the inner loop employs a multivariable version of super-twisting algorithm (MSTA) in place of high-gain position control, providing robustness against unknown environmental stiffness. The contributions of this manuscript are as follows:
\begin{itemize}
  \item Different from previous works, the proposed set-valued admittance control is with two nonsmooth functions in a coupled manner by considering both the interaction safety and coupled nature of manipulators. 
  \item To make the set-valued admittance controller implementable in digital platforms, this paper also derives a framework of discretization for the out and inner set-valued loops such that different discretization methods.
  \item Simulations and experiments have validated the proposed admittance control by showing the preventing from snapping backing after impacting to unknown and changed environments in terms of stiffness.
\end{itemize}


\par

The rest of this paper is organized as follows: Section \ref{sec:proposed} introduces a new admittance controller featuring two set-valued loops and presents its discrete-time implementation algorithm. Section \ref{Sec:impact} validates the proposed controller through simulations and experiments. Finally, Section \ref{sec:conclusion} summarizes the paper and discusses future work.

\section{Impact-Contact Admittance Control}
\label{sec:proposed}

Let us consider one manipulator with multiple degrees of freedom (DoF) that experiences contact force when interacting with its surroundings:
\begin{flalign}\label{equ:robot}
{M}({q})\ddot{{q}}+C(q,\dot{q})\dot{q}+G(q)=\tau+f_{c}+f_e
\end{flalign}
where $q\in \mathbb{R}^n$ is the position, $M(q)\in \mathbb{R}^{n\times n}$ and $C(q,\dot{q})\in  \mathbb{R}^{n\times n}$ are the inertia and Coriolis matrixes of the controlled robot, $G(q)$ represents the gravitational force, $f_c=J^T \bar{f}_c \in \mathbb{R}^n$ is the torque
caused by the contact force $\bar{f}_c\in \mathbb{R}^6$ in the Cartesian space acting from its environments and measured by equipped 6-axis force sensor, $J\in \mathbb{R}^{6\times n}$ is the Jacobian, $\tau \in \mathbb{R}^n$ is the actuation input by following a law of admittance control, and $f_e=J^T\bar{f}_e$ with $\bar{f}_e$ unable to measure by the force sensor, which accounts for the external force disturbances and unmodeled dynamics. Here, it is assumed that only $q$, $\dot{q}$, $\ddot{q}$ and $f_c$ are available and the objective is to design the law of admittance control such that the interaction between the above system and its unknown environments is safely subjected to the actuation saturation or limited torque.

\begin{figure}[!t]
	\centering
		\includegraphics[width=0.45\textwidth]{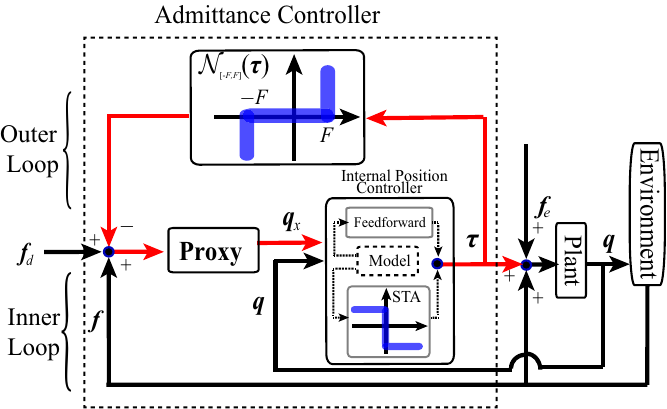}
	\caption{Illustration of the proposed admittance controller with an inner loop of MSTA (Multivariable Sliding Twisting Algorithm) and an outer loop of normal cone $\mathcal{N}_{\mathcal{F}}$ that is equivalent to the inversed map of a first-order SMC.}
	\label{Fig_Proxy_system}
\end{figure}

The manipulator \eqref{equ:robot} is typically controlled by an admittance system that includes a proxy system when interacting with its surroundings.
When contacting with environments of unknown stiffness, the potential unsafe behaviors of conventional admittance control systems can be analyzed as follows.
If the environment's stiffness does not match the parameters of the admittance controller—for instance, if the stiffness is high and ${f}_c$ represents an impact contact force—this causes a rapid change in the proxy system's position $q_x$. Consequently, there is a sharp increase in position tracking error denoted as $q_e=q_x-q$. The high-gain inner position control in conventional admittance controllers responds to this large error by producing significant torque $\tau$ until actuation saturation. This can lead to unsafe windup behaviors such as overshooting and oscillations during interaction. To mitigate these unsafe behaviors, a set-valued loop constrains the actuation torque $\tau$, producing anti-windup effects that prevent these issues.
Additionally, if there is an unmeasured and unknown disturbance force represented by $f_e$, it causes the actual position $q$ to quickly deviate from $q_x$ of the proxy system, also resulting in large position tracking error and further unsafe behaviors. Another set-valued loop is required to reject this unmeasured disturbance force $f_e$, enhancing the interaction safety.

\subsection{Proposed Set-Valued Admittance Controller }

To combine previous strategies' advantages of admittance control while addressing the above safety concerns, we propose an admittance control strategy featuring two set-valued loops, an outer set-valued loop using multivalued sliding mode control (SMC) and an inner set-valued loop employing a multivariable super-twisting algorithm (MSTA) \cite{NAGESH2014984,Moreno_2022_MSTA}:
\begin{subequations}\label{equ:kikuuwe_adm_proposed}
\begin{flalign}
& \tau\in -F\mathrm{Sgn}( M_{x} \ddot{q}_{x}+B_{x} \dot{q}_{x} -  f_c-f_{d}) \label{equ:kikuuwe_adm_proposeda} \\
& \tau= M(q) \ddot q_r+ C(q,\dot{q})\dot{ q}_r+G(q)+k_1s+M(q)u_s \label{equ:kikuuwe_adm_proposedb} \\
&  u_s\in k_2 \frac{s}{\|s\|^{1/2}}+v, \,\dot{v}\in k_3\frac{s}{\|s\|}+k_4 s \label{equ:kikuuwe_adm_proposedc}
\end{flalign}
\end{subequations}
where the symbol ``$\in$" is employed here instead of the equality ``$=$" because the righ hand sides of the corresponding equations are sets, and it can be found in the literature \cite{Brogliato_2017_TAC,Xiong_2022_TSCII}. The variable $q_x\in \mathbb{R}^n$ represents the position of a proxy system, while $M_{x}\in \mathbb{R}^{n\times n} $ and $B_{x}\in \mathbb{R}^{n\times n}$ are the designed invertible matrices for the proxy system. Each entry $F_i>0$ in the diagonal matrix $F=\diag(F_1,\cdots, F_n)$ represents the torque limit for the $i$th joint, constrained by safety or physical driving torque considerations, $i=1,2,\cdots, n$. Here, $f_d\in \mathbb{R}^n$ is the desired torque, and $  q_{e}:=q_{x}-q$ is the tracking error between the actual position $q$ and the proxy position $q_x$. The reference velocity is defined as $\dot{q}_r :=\dot{q}_{x}+\Lambda q_{e}$ while $s= \dot{q}_r-\dot{q}=\dot{q}_{e}+\Lambda q_{e}$ is the sliding mode variable where $\Lambda, k_j>0,j=2,\cdots,4$ are positive constants, and $k_1\in\mathbb{R}^{n\times n}$. For any vector $x\in \mathbb{R}^n$, the function $\mathrm{Sgn}( x):=[\sgn(x_1),\sgn(x_2),\cdots, \sgn(x_n)]^T$ in \eqref{equ:kikuuwe_adm_proposed} is element-wised defined, where $\sgn(x_i), i\in\{1,2,\cdots, n\}$ is the set-valued signum function. The two signum definitions are distinguished: $\forall z\in \mathbb{R}$,
\begin{flalign}\label{equ:sgn}
& \mathrm{sign}(z)\!\!=\!\! \left\{\begin{array}{ll} \!\!\! \left\{\! \dfrac{z}{|z|}\!\right\} \hspace{-0.5em} &\mbox{if }z\!\!\neq\!\! 0 \\ \!\!\! 0  \hspace{-0.5em} &\mbox{if }z\!=\!\! 0 \end{array}\right.  \mathrm{sgn}(z)\!\!=\!\!  \left\{\begin{array}{ll}\!\!\!  \left\{\! \dfrac{z}{|z|}\!\right\} \hspace{-0.5em} &\mbox{if }z\!\!\neq\!\! 0 \\ \!\!\! [-1,1]  \hspace{-0.5em} &\mbox{if }z\!\!=\!\! 0, \end{array}\right. &
\end{flalign}
 where the set-valued version $\sgn(\cdot)$ can be found in the fields of sliding mode control (SMC) \cite{Acary_2012_Euler,Brogliato_2020_STA}. When $n=1$, the MSTA \eqref{equ:kikuuwe_adm_proposedc} reverts to the super-twisting algorithm (STA). Please note that the literature presents various versions of MSTA (see \cite{NAGESH2014984,Moreno_2022_MSTA,Ding_TCYB_2023,Hsu_2024_TAC} and references therein). Consequently, \eqref{equ:kikuuwe_adm_proposedc} can be expressed in different forms depending on the version of MSTA used.

As shown in Fig.~\ref{Fig_Proxy_system}, \eqref{equ:kikuuwe_adm_proposeda} represents the constraint of torque $\tau$ in the outer loop and   \eqref{equ:kikuuwe_adm_proposedb}-\eqref{equ:kikuuwe_adm_proposedc} represent the inner loop control. 
In the inner loop control, \eqref{equ:kikuuwe_adm_proposedb} adheres to the conventional robust control method for manipulators. This includes a nonlinear feedforward term $M(q) \dot q_r + C(q,\dot{q}) q_r + g(q,\dot{q})$ and a linear feedback control $k_1s$. Additionally, it also incorporates the set-valued feedback control $u_s$ to ensure that $q$ approaches $q_x$, even when in contact with unknown environments in term of stiffness.
One can also observe that there are two set-valued inclusions, ``$\in$", as shown in \eqref{equ:kikuuwe_adm_proposeda} and \eqref{equ:kikuuwe_adm_proposedc}, respectively.
The inclusion \eqref{equ:kikuuwe_adm_proposedc} about $u_s$ is the robust control with the set-valued MSTA with positive gains $k_2, k_3>0$ and $k_4\geq 0$. 
The inclusion \eqref{equ:kikuuwe_adm_proposeda} represents a first-order multivalued set-valued SMC as in \cite{Brogliato_2017_TAC,MOJALLIZADEH2024140} but with the sliding surface of proxy system $M_{x} \ddot{q}_{x}+B_{x} \dot{q}_{x} -  f_c-f_{d}$. It is inspired by the concept of the normal cone $ M_{x} \ddot{q}_{x}+B_{x} \dot{q}_{x} \in f_c+f_d-\mathcal{N}_{\mathcal{F}}(F^{-1}\tau)$ where the torque $\tau$ is constrained by the normal cone $\mathcal{N}_\calF(F^{-1}\tau)$, defined as $\mathcal{N}_{\mathcal{F}}(y)=\{z\in \mathbb{R}^n| z(y^\ast-y)\leq 0, \forall y^\ast \in \calF\}$ if $y\in \calF$ and $\mathcal{N}_{\mathcal{F}}(y)=\emptyset$ if $y \notin \calF$, as shown in Fig.~\ref{Fig_Proxy_system}. The normal cone $\mathcal{N}_{\mathcal{F}}(F^{-1}\tau)$ constrains the torque $F^{-1}\tau$ within the set $\mathcal{F}$ and it has the equivalent transformation $ F^{-1}\tau \in \operatorname{Sgn}(y) \Longleftrightarrow \mathcal{N}_{\mathcal{F}}(F^{-1}\tau ) \ni y$ with $\mathcal{F}=[-1, 1]^{n}$. The MSTA in \eqref{equ:kikuuwe_adm_proposedc} with the intermediate variable $v$ can be used to reject differentiable disturbances, i.e., the disturbances matched with the jerk \cite{NAGESH2014984,Moreno_2022_MSTA}.

The approach using two set-valued operators in \eqref{equ:kikuuwe_adm_proposeda} and \eqref{equ:kikuuwe_adm_proposedc} not only keeps the actuation torque $\tau_i$ of each joint within the safe range $[-F_i, +F_i]$, $i= 1,\cdots, n$, but also prevents excessive deviation between the proxy position $q_x$ and $q$. This helps mitigate safety issues like oscillations, snapping back, and overshooting that typically arise from significant errors between $q_x$ and $q$, denoted as $q_e=q_x-q$. Such errors can be caused by external disturbance forces ($f_e$, which causes rapid changes in $q$) or large contact forces ($f_c$, which causes rapid changes in $q_x$). These situations often occur when nearby humans inadvertently affect the robot's body position instead of its end-effector equipped with force sensors.
In \eqref{equ:kikuuwe_adm_proposed}, when the manipulator model terms $M(q)$, $C(q,\dot{q})$, and $G(q)$ are unavailable, their estimates $\hat M(q)$, $\hat C(q,\dot{q})$, and $\hat G(q)$ can be used and then $f_e$ includes the model uncertainties $\Delta M(q)=M(q)-\hat{M}(q)$, $\Delta C(q,\dot{q})=C(q,\dot{q})-\hat{C}(q, \dot{q})$, and $\Delta(q)= G(q)-\hat{G}(q)$, as demonstrated in \cite{Brogliato_2017_TAC}. The set-valued SMC and MSTA still ensure that $q$ approaches $q_x$ due to their robustness.
On the other hand, $u_s $ enhances the conventional manipulator controller $ M(q) \ddot{q}_r + C(q,\dot{q})\dot{q}_r + G(q) + k_1 s$ by leveraging the differentiability of impact-contact $ f_c = J^T\bar{f}_c $. This ensures quick convergence of $ q_x $ towards $ q $, following any deviation caused by the contact force $ f_c $. Since both the measurability and differentiability are available for $ f_c $, the gain matrixes $k_j$, $ j\in\{1,2,3,4\}$ can be adjusted in real-time based on the magnitude of $ \dot{f}_c$, using filtering differentiator techniques with $f_c$ as the input \cite{Hanan_2022_TAC,Brogliator_2023_CEP}.

\par

\subsection{Stability Analysis}
The closed-loop system is interconnected by the system \eqref{equ:robot} and controller \eqref{equ:kikuuwe_adm_proposed}. Due to the equivalent transformation $\forall y \in \mathbb{R}^n, F^{-1}\tau \in \operatorname{Sgn}(y) \Longleftrightarrow \mathcal{N}_{\mathcal{F}}(F^{-1}\tau ) \ni y$ where ``$\ni$" can be found in \cite[Eq. (2.81)]{Acary_2008_book} and \cite[Eq. (2.139)]{Acary_2010_book}, the proposed admittance control can be equivalently written as follows:
\begin{subequations}\label{equ:stab}
\begin{flalign}
& M_{x} \ddot{q}_{x}+B_{x} \dot{q}_{x} =  f_c+f_{d}+\lambda \label{equ:staba} \\
& \tau= M(q) \ddot q_r+ C(q,\dot{q})\dot{ q}_r+G(q)+k_1s+M(q)u_s  \label{equ:stabb} \\
&  u_s\in \!k_2 \frac{s}{\|s\|^{1/2}}+v, \,\dot{v}\in k_3\frac{s}{\|s\|}+k_4 s \label{equ:stabc}
\end{flalign}
\end{subequations}
with a nonsmooth map $\lambda\in  -\mathcal{N}_{\mathcal{F}}(F^{-1}\tau)$
where $\lambda \in \mathbb{R}^n$ is an intermediate variable. Due to the normal cone property, $\lambda^T \tau\leq 0$, which means that the stability of subsystem \eqref{equ:robot}\eqref{equ:stab} with contact model of environments is more important than that of the entire system. The latter consists of a nonsmooth map $\lambda\in -\mathcal{N}_{\mathcal{F}}(\tau)$, a contact model and closed-loop system \eqref{equ:robot}\eqref{equ:stab} \cite{Kikuuwe_TRO_2019}. Due to the passivity property of the normal cone, we can only consider the case $\lambda=0$. Then, let us check the two cases separately: $f_c=0$ and $f_c\neq 0$.




\textbf{1) Contact force} $f_c=0$: it means that there is no contact between the system \eqref{equ:robot} and its surrounding environments. In this scenario, one can view the system \eqref{equ:robot} as being perturbed by $f_e$ and controlled by \eqref{equ:stabc}, and it does not make sense for \eqref{equ:staba} by setting $f_d=0$. By substituting \eqref{equ:stabc} into \eqref{equ:robot}, the error dynamics is as follows:
\begin{flalign} \label{equ:error_dy}
&M(q)\dot{s}+ C(q,\dot{q})s \in -f_e-k_1s-M(q)u_s.
\end{flalign}
Let us assume that the unknown disturbance $ f_e$ and its derivative are bounded, i.e., $\|f_e\| \leq \delta_1 $ and $\|\dot{f}_e\| \leq \delta_2 $ with some constant $\delta_1,\delta_2>0$, and $k_1\succ 0$. Let us consider the Lyapunov function as $V(s)=s^TM(q)s/2$ and one has:
\begin{flalign}
& \dot{V}(s)=s^T\frac{1}{2}\dot{M}(q)s+s^T(-C(q,\dot{q})s-f_e-(k_1s+u_s))\nonumber & \\
&  =\!-s^Tk_1s\!-s^Tf_e\!-s^TM(q)u_s\leq\! -\kappa_2 \|s\|^2+\|s\| \|f_e\| \nonumber &\\
&\leq \ -\|s\|^2 \gamma_{1,\min}+\!\|s\|\delta_1\leq-\gamma_{1,\min}\frac{2V}{\lambda_{\max}(M)}\!  \nonumber &\\
&+\delta_1\sqrt{\frac{2V}{\lambda_{\min}(M)}} 
\end{flalign}
where $\lambda_{\min} (M)\|s\|^2 \leq V(s)\leq \lambda_{\max}(M) \|s\|^2$ with $\lambda_{\min}$ and $\lambda_{\max}$ representing the minimum and maximum eigenvalues of $M(q)$, respectively, $\gamma_{1,\min}=\lambda_{\min} (k_1)>0$, the skew-symmetry of $\dot{M}(q)-2C(q,\dot{q})$, and $s^TM(q)u_s\geq 0$, $M(q)\succeq 0$, $\forall v_0=v(0)\geq 0$ have been employed. 

Let $\bar V(s)=\sqrt{V(s)}$, and then one has $\dot{\bar V}=\dot{V}/(2\sqrt{V})$ and the following expression:
\begin{flalign}\label{equ:compar}
& \dot{\bar V}\leq  -\dfrac{\gamma_{1,\min}}{\lambda_{\max}(M)}\bar V+\delta_1\sqrt{\dfrac{1}{2\lambda_{\min}(M)}}. 
\end{flalign}
With the Comparison Lemma\cite{khalil_nonlinear_2002}, this inequality \eqref{equ:compar} leads to the finite-gain $\mathcal{L}_p$ stability and the input-to-state stability (ISS)\cite{Guanya_2019_ICRA}. Furthermore, the sliding variable $s$ and the tracking error $q_e$ converge exponentially to their respective error bounds: $\lim_{t\to \infty}\|s\|=\delta_1\lambda_{\max}(M)/(\lambda_{\min}(M)\gamma_{1,\min})$ and $\lim_{t\to \infty}\|q_e\|=\delta_1\lambda_{\max}(M)/(\lambda_{\min}(M)\gamma_{1,\min}\Lambda)$, according to the definition $s= \dot{q}_r-\dot{q}=\dot{q}_{e}+\Lambda q_{e}$.

\textbf{2) Contact force} 
$f_c \neq 0$: it indicates a contact between the system \eqref{equ:robot} and its environment. For simplicity, assume an elastic contact with the surroundings, that is, $\tau_c$ is differentiable, although more complex contact models, such as the spring-damping model, are also possible.
 Because of the equivalent transformation $ F^{-1}\tau \in \operatorname{Sgn}(y) \Longleftrightarrow \mathcal{N}_{\mathcal{F}}(F^{-1}\tau ) \ni y$, from \eqref{equ:kikuuwe_adm_proposeda}, $\lambda=0$ means the sliding variable $M_{x} \ddot{q}_{x}+B_{x} \dot{q}_{x}- f_c-f_{d}=0$ and $F^{-1}\tau \in [-1,+1]^n$, and $q_x, \dot{q}_x$ are all bounded.
By substituting \eqref{equ:kikuuwe_adm_proposedb} and \eqref{equ:kikuuwe_adm_proposedc} into \eqref{equ:robot}, the error dynamics is as follows:
\begin{subequations} \label{equ:error_dy02}
\begin{flalign}
&  M(q)\dot{s}+C(q,\dot{q})s\in -f_c-f_e-k_1s-M(q)u_s \label{equ:error_dy02a} \\
& u_s\in \!k_2 \frac{s}{\|s\|^{1/2}}+v, \,\dot{v}\in k_3\frac{s}{\|s\|}+k_4 s. \label{equ:error_dy02b}
\end{flalign}
\end{subequations}
In the case $f_c\neq 0$, for simplicity, let us select $k_1=-C(q,\dot{q})+\gamma_1 M(q)$, and $\gamma_1>0$ such that $k_1\succeq 0$, and then \eqref{equ:error_dy02} can be further equivalently rewritten as the form like the MSTA:
\begin{subequations} \label{equ:error_dy03}
\begin{flalign}
&  \dot{s}_1\!\in \!-(\!\kappa_1s_1+\!k_2 s_1{\|s\|^{-\frac{1}{2}}})+s_2 \\
&   \dot{s}_2\in - k_3s_1{\|s_1\|}^{-1}-k_4s_1-\Delta, 
\end{flalign}
\end{subequations}
where $s_1:=s$, $\Delta:=\frac{\mathrm{d}}{\mathrm{d}t}[ M^{-1}(q)(f_c+f_e)]$, $\kappa_1=\gamma_1$, and $s_2:=-v-M^{-1}(q)(f_c+f_e)$. Let the term $\Delta$ be bounded, satisfying $\|\Delta\| \leq \delta_3$ for some constants $\delta_3 > 0$. 
Then, \eqref{equ:error_dy03} becomes the standard formulation of the MSTA as shown in \cite{NAGESH2014984, Moreno_2022_MSTA}. There exist appropriate gains $\kappa_1$, $\kappa_2$, $k_3$, and $k_4$ such that $s_1$ and $\dot{s}$ converge to zero in finite time and remain at zero thereafter. The corresponding proof can be provided by designing a Lyapunov function using MSTA techniques from existing literature \cite[Eq.(12)]{NAGESH2014984} or \cite{Moreno_2022_MSTA}, then combining it with \eqref{equ:error_dy03}.

From the above analysis, one can see that for both $f_c=0$ and $f_c\neq 0$, the error dynamics \eqref{equ:error_dy} and \eqref{equ:error_dy03} about $s$ and $q_e$ show the stability. If the stiffness of the environment is high, the term $\dot{f}_c$ represents an impact contact force that causes rapid changes in $q_x$ within the proxy system. As a result, there is a rapid increase in positional tracking error denoted as $q_e=q_x-q$. The conventional admittance controller responds to this error by producing large torque $\tau$, which can lead to unsafe behaviors such as overshooting and oscillations in system \eqref{equ:robot}. In contrast, our proposed first set-valued SMC on $\tau$ in equation  \eqref {equ:kikuuwe_adm_proposeda} not only constrains torque $F^{-1}\tau\in[-1,+1]^n$, but also prevents excessive deviation of $q_x$ from $q$ caused by the impact force.

\subsection{Discrete-Time Implementation of Admittance Controller }

Due to its set-valued characteristics, the proposed admittance controller \eqref{equ:kikuuwe_adm_proposed} is represented as a differential inclusion (DI) using the symbol “$\in$,” making direct implementation impossible. Conventional explicit Euler discretization methods compromise the set-valued nature, while implicit Euler methods maintain this set-valued aspect, as shown in\cite{Brogliato_2017_TAC} and\cite{andritsch2023modified}. However, \eqref{equ:kikuuwe_adm_proposed} incorporates two interdependent set-valued nonsmooth operators: the first-order set-valued SMC and MSTA. 
This approach significantly differs from those in\cite{Brogliato_2017_TAC} and\cite{andritsch2023modified}, which involve only one nonsmooth operator—either the first-order set-valued SMC or MSTA. The interdependence of these two operators makes it challenging to discretize \eqref{equ:kikuuwe_adm_proposed} using conventional implicit Euler methods.

\par
\subsubsection{Framework of discretization of \eqref{equ:kikuuwe_adm_proposed}}

The inclusion \eqref{equ:kikuuwe_adm_proposeda} can be equivalently written as follows:
\begin{flalign}\label{equ:discrete_1}
 \tau & \in -F \operatorname{Sgn}\left(\ddot{q}_x+{M_x}^{-1}({B_x} \dot{q}_x-{\left(f_c+f_d\right)})\right)
\end{flalign}
where the fact $ \operatorname{Sgn}(x)=\operatorname{Sgn}(M_xx), \forall x\in \mathbb{R}^n, M_x>0$ was used. Its implicit Euler discretization method is as follows:
\begin{subequations}\label{equ:tau01}
\begin{flalign}
& \tau_k \in-F \operatorname{Sgn}\left(q_{x,k}-q_{x,k}^\ast\right) \label{equ:tau01a} \\
& u_{x,k}^\ast \triangleq {(M_x+B_x h)}^{-1}({M_x \dot{q}_{x,k-1}+h\left(f_{c,k}+f_{d,k}\right)}) \\
& q_{x,k}^\ast \triangleq q_{x,k-1}+h u_{x,k}^\ast \\
&  \dot{q}_{x,k}=(q_{x,k}-q_{x,k-1})/h,
\end{flalign}
\end{subequations}
where $h$ is the size of time-stepping, i.e., $h=t_{k+1}-t_k$,
and $k\geq 0$ is the index of time-stepping with $t_k=kh$, $q_{x,k}=q_x(t_k)$, $f_{c,k}=f_c(t_k)$ and so on. The implicit Euler discretization of \eqref{equ:kikuuwe_adm_proposedb} is as follows:
\begin{subequations}\label{equ:tau02}
\begin{flalign}
& \tau_k =(M_k/h^2+\hat{K})(q_{x,k}-q_{1,k}^\ast)  \label{equ:tau02a} \\
& q_{1,k}^\ast \triangleq q_k+(M_k/h^2+\hat{K})^{-1}({\phi_{b,k}-\phi_{a,k}}) \label{equ:tau02b}\\
& \phi_{a,k} \triangleq \frac{(M_k+{C_k}h)q_k+Bhq_{k-1}}{h^2}+G_k+M_ku_{s,k} \label{equ:tau02c} \\
&  \phi_{b,k} \triangleq \frac{M_k(q_{x,k-1}+hu_{x,k-1})}{h^2}+\frac{\hat{B}q_{x,k-1}}{h}, \label{equ:tau02d}
\end{flalign}
\end{subequations}
where $\hat{K}=\hat{B}/h+K$, $\hat{B}=B+C_k$, $B=M_k\Lambda+k_1I_n$, $K=(C_k+k_1I_n)\Lambda$, $C_k=C(q_k,\dot{q}_k)$, $M_k=M(q_k)$ and so on.

 Then, with the help of transformation $ F^{-1}\tau \in \operatorname{Sgn}(y) \Longleftrightarrow \mathcal{N}_{\mathcal{F}}(F^{-1}\tau ) \ni y$, from \eqref{equ:tau01a} and \eqref{equ:tau02a}, one can obtain the equivalent expression:
\begin{subequations}\label{equ:equ}
\begin{flalign}
&\tau_k-\tau_k^\ast \in -\mathcal{N}_\mathcal{F}(F^{-1}\tau_k), \\
&\tau_k^\ast=(M_k/h^2+\hat{K})(q_{x,k}^\ast-q_{1,k}^\ast).
\end{flalign}
\end{subequations}
The following equivalence can be applied to \eqref{equ:equ}: $\forall x,y\in \mathbb{R}^n$, a closed
convex non-empty set $\mathcal{C}\subseteq \mathbb{R}^n$, and a weighted inner product $ \langle \cdot, \cdot \rangle_W$ with a matrix $W=W^T\in \mathbb{R}^{n\times n}$ \cite[Proposition 2.37]{Acary_2010_book},
\begin{flalign} \label{equ:equivalence}
W(x-y)\in -\mathcal{N}_{\mathcal{C}}(x) \Longleftrightarrow x =\operatorname{Proj}_{W}(\mathcal{C}; y),
\end{flalign}
where $\operatorname{Proj}_{W}(\mathcal{C}; y)=\arg\min\frac{1}{2}\langle z-y , z-y \rangle_W+\Psi_\mathcal{C}(z)$ with the indicator function $\Psi_\mathcal{C}(z)$. With the help of \eqref{equ:equivalence}, \eqref{equ:equ} can be equivalently transformed into:
\begin{flalign}\label{equ:torq}
&\tau_k=F\operatorname{Proj}(\mathcal{F}; F^{-1}\tau_k^\ast)
\end{flalign}
where $\operatorname{Proj}_W(; )$ is simply rewritten as $\operatorname{Proj}(; )$ when $W=I_n$ is the identity matrix. Due to $\mathcal{F}=[-1,+1]^n$ and $F=\mathrm{diag}(F_1, F_2, \cdots, F_n)$, the $i$th entry of $\tau_k$ in \eqref{equ:torq} can be calculated as 
\begin{flalign}\label{equ:elem}
(\tau_k)_{i}=F_i\min\Bigl\{\frac{|(\tau_k^\ast)_i|}{F_i}, 1 \Bigr\}\mathrm{sign}\Bigl((\tau_k^{\ast})_i\Bigr)
\end{flalign}
where $\mathrm{sign}()$ is defined in \eqref{equ:sgn}. In the scalar case, i.e., $n=1$, \eqref{equ:torq} can be written as $\tau_k=\operatorname{Proj}([-F_1, F_1];\tau_k^\ast)$.


The computation procedure from \eqref{equ:tau01} to \eqref{equ:elem} for $\tau_k$ establishes a framework for digitally implementing \eqref{equ:kikuuwe_adm_proposed} with two interdependent set-valued operators: the first-order SMC and the MSTA, based on implicit Euler discretization and characteristics of the proxy system. Unlike the work in\cite{Brogliato_2017_TAC}, this procedure computes $\tau_k$ without iterations, which is beneficial for real-time control (see \cite[Remark 5 and Sec. VII]{Brogliato_2017_TAC}). Additionally, it separates the two interdependent set-valued operators—first-order SMC and MSTA—allowing for the reuse of various conventional discretizations of MSTA.

\subsubsection{Various Implementations of MSTA} The MSTA \eqref{equ:kikuuwe_adm_proposedc} can be discretized by various methods. 
The conventional discretization of MSTA is based on the explicit Euler method:
\begin{subequations} 
\begin{flalign}  \label{equ:MSTA_Explicit}
&u_{s,k}\!=\!v_k\!+\!k_2\frac{s_k}{\|s_k\|^{\frac{1}{2}}}, \\
& v_{k+1}\!=\!v_k+hk_3\frac{s_k}{\|s_k\|}+k_4s_k. 
\end{flalign}
\end{subequations}
As pointed out in our previous research \cite{Xiong_2023_MSTA}, this explicit Euler method also suffers from the numerical chattering.

The literature presents various discretization methods for the conventional super-twisting algorithm (STA), i.e., single-variable STA, including explicit Euler, implicit Euler, and semi-implicit Euler methods\cite{Xiong_2019_TCASII,Xiong_2022_TCASII_STA_Semi,andritsch2023modified,Brogliato_2020_STA,KOCH_2019_STA,Levant_2022_TAC}. However, these discretization algorithms can only be applied when $n=1$, where the MSTA in equation \eqref{equ:kikuuwe_adm_proposedc} becomes equivalent to the STA. For $n \neq 1$, a careful redesign of the implicit Euler discretization in equation \eqref{equ:kikuuwe_adm_proposedc} is required.

To implement the MSTA \eqref{equ:kikuuwe_adm_proposedc} without suffering from the numerical chattering, let us discretize \eqref{equ:error_dy03} with an implicit Euler method:
\begin{subequations} \label{equ:proposed_admin_002}
\begin{flalign} 
& s_{1,k+1}=s_{1,k}-hk_2{m}_{1,k}+h{s}_{2,k+1} \label{equ:proposed_admin_002a} \\
& s_{2,k+1}=s_{2,k}-hk_3 m_{2,k}-h\Delta_k        \label{equ:proposed_admin_002b} 
\end{flalign}
\end{subequations}
where ${m}_{1,k}\in \partial \Psi_1({s}_{1,k+1})$, $m_{2,k}\in \partial \Psi_2({s}_{1,k+1})$, $\partial \Psi_i$ represents the subdifferential of $\Psi_i$, $\Psi_1(\cdot)=\frac{2}{3}\|\cdot\|^{3/2}+\frac{\alpha_1}{2}\|\cdot\|^2$, $\Psi_2(\cdot)=\|\cdot\|+\frac{\alpha_2}{2}\|\cdot\|^2$, and $\alpha_1=\kappa_1/k_2$ and $\alpha_2=k_4/k_3$. The subdifferential of a convex, proper, lower-semicontinuous functions $f$ is defined as follows: 
\begin{flalign}\label{equ:subdiff}
\partial f(v)=&\left\{w \in \mathbb{R}^n \mid\langle w, s-v\rangle \leq f(s)-f(v)\right. \nonumber \\
&\forall \left.s \in \mathbb{R}^n\right\}.
\end{flalign}

One can see that $m_{i,k}$, $i\in\{1,2\}$, depends on the unknown state $s_{1,k+1}$. By substituting \eqref{equ:proposed_admin_002b} into \eqref{equ:proposed_admin_002b}, one has 
\begin{flalign}\label{equ:s_1,k}
s_{1,k+1}=s_{1,k}-hk_2{m}_{1,k}-h^2k_3 m_{2,k}+hs_{2,k}+h^2\Delta_k
\end{flalign}
which contains the unknown term $h^2\Delta_k$ and $hs_{2,k}$ (see the definition of $s_2$ in \eqref{equ:error_dy03}).
To make $m_{i,k}$ implementable, let us rewrite \eqref{equ:proposed_admin_002} as follows:
\begin{subequations} \label{equ:proposed_admin_003}
\begin{flalign} 
& s_{1,k+1}=\hat{s}_{1,k+1}+h{s}_{2,k+1} \label{equ:proposed_admin_003a} \\
& s_{2,k+1}=s_{2,k}-hk_3 \hat{m}_{2,k}+h\Delta_k        \label{equ:proposed_admin_003b} \\
& \hat{s}_{1,k+1}=s_{1,k}-hk_2\hat{m}_{1,k}-h^2k_3\hat{m}_{2,k}\label{equ:proposed_admin_003c}
\end{flalign}
\end{subequations}
where $\hat{m}_{1,k}\in \partial \Psi_1(\hat{s}_{1,k+1})$ and $\hat{m}_{2,k}\in \partial \Psi_2(\hat{s}_{1,k+1})$ are approximations of ${m}_{1,k}$ and ${m}_{2,k}$, respectively, while $\hat{s}_{1,k+1}$ in \eqref{equ:proposed_admin_003c} represents the nominal state of ${s}_{1,k+1}$ in \eqref{equ:s_1,k} by considering both $hs_{2,k}$ and $h^2\Delta_k$ as unknown disturbances, as done in our previous work \cite{Xiong_2022_TCASII_STA_Semi}.

\subsubsection{Unperturbed Case}
 The stability of discrete-time system \eqref{equ:proposed_admin_003}, which is the implicit Euler discretization of closed-loop system \eqref{equ:error_dy03}, under the bounded derivatives $\|\Delta_k\|\leq \delta_3$ can be analysed.
Let us consider the following discrete-time Lyapunov function: $V_{1,k}=k_3\Psi_2(\hat{s}_{1,k})+\|s_{2,k}\|^2_2/2 $. In the case of $\Delta_k\equiv 0$, one has:
\begin{flalign}\label{equ:Lyap}
&\Delta V_{1,k}=V_{1,k+1}-V_{1,k}\leq k_3\left\langle \hat{m}_{2, k}, \hat{s}_{1, k+1}-\hat{s}_{1, k}\right\rangle \nonumber \\
&+\left\langle s_{2, k+1}, s_{2, k}-h k_3 \hat{m}_{2, k}\right\rangle- \frac{1}{2}\left\|s_{2, k+1}\right\|_2^2-\frac{1}{2}\left\|s_{2, k}\right\|_2^2 \nonumber \\
&=\left\langle h k_3 \hat{m}_{2, k}, s_{2, k+1}-k_2 \hat{m}_{1, k}\right\rangle-h k_3\left\langle \hat{m}_{2, k}, s_{2, k+1}\right\rangle  \nonumber \\
& \quad- \frac{1}{2}\left\|s_{2, k+1}-s_{2, k}\right\|_2^2 \nonumber \\
&=-h k_2 k_3\left\langle \hat{m}_{1, k}, \hat{m}_{2, k}\right\rangle-\frac{1}{2}\left\|s_{2, k+1}-s_{2, k}\right\|_2^2=-S_k 
\end{flalign}
from which one has $S_k\geq 0, \forall k\in \mathbb{N}$ and $V_{1,k+1}-V_{1,k}\leq -S_k$. Therefore, as $k\to \infty$, one has $S_k\to 0$, because $V_{1,k}$ is decreasing. Furthermore, as $S_k\to 0$,  one has $s_{2, k+1}-s_{2, k}\to 0$ from \eqref{equ:Lyap}; $\hat{m}_{2,k} \to 0$ from \eqref{equ:proposed_admin_003b} and the assumption $\Delta_k\equiv 0$; $\hat{s}_{1, k+1} \to 0$ and $\hat{m}_{1,k}\to 0$ due to $0\in \partial \Psi_2(0)$. From the above analysis, one has the conclusion that when there is no contact and no disturbance, i.e., $\Delta_k\equiv 0$, the asymptotic stability of the origin for \eqref{equ:proposed_admin_003} can be achieved.

 

\subsubsection{Perturbed Case} 
When $\Delta_k\neq 0$, as done in our previous work\cite{Xiong_2022_TCASII_STA_Semi}, let us rewrite \eqref{equ:proposed_admin_003} as follows:
\begin{subequations} \label{equ:proposed_admin_004}
\begin{flalign} 
& s_{1,k+1}=\hat{s}_{1,k+1}+h\hat{s}_{2,k+1}+h^2\Delta_k \label{equ:proposed_admin_004a} \\
& s_{2,k+1}=\hat{s}_{2,k+1}+h\Delta_k        \label{equ:proposed_admin_004b} \\
& \hat{s}_{1,k+1}=s_{1,k}-hk_2\hat{m}_{1,k}-h^2k_3\hat{m}_{2,k}\label{equ:proposed_admin_004c}\\
& \hat{s}_{2,k+1}=s_{2,k}-hk_3 \hat{m}_{2,k}  \label{equ:proposed_admin_004d}
\end{flalign}
\end{subequations}
where $\hat{s}_{1,k+1}$ and $\hat{s}_{2,k+1}$ are the nominal states of $s_{1,k+1}$ and $s_{2,k+1}$, respectively, and they can be further rewritten as follows:
\begin{subequations} \label{equ:hats2}
\begin{flalign} 
& \hat{s}_{1,k+1}=\hat{s}_{1,k}+h\hat{s}_{2,k+1}-hk_2\hat{m}_{1,k} \label{equ:hats2a}\\
& \hat{s}_{2,k+1}=\hat{s}_{2,k}-hk_3 \hat{m}_{2,k}+h \Delta_{k-1}.  \label{equ:hats2b}
\end{flalign}
\end{subequations}
 Let us consider the following discrete-time Lyapunov function: $V_{2,k}=k_3\Psi_2(\hat{s}_{1,k})+\|\hat{s}_{2,k}-k_2 \hat{m}_{1,k-1}\|^2_2/2 $, $\Delta V_{2,k}= V_{k+1}-V_k$, and from \eqref{equ:hats2}, one has:
\begin{flalign}\label{equ:Lyap02}
 \Delta V_{2,k} &\leq-\frac{1}{2}\left\|\hat{s}_{2, k+1}-k_2 \hat{m}_{1, k}-\left(\hat{s}_{2, k}-k_2 \hat{m}_{1, k-1}\right)\right\|_2^2 \nonumber \\
- & \frac{k_2}{h}\left\langle\hat{m}_{1, k}-\hat{m}_{1, k-1}, \hat{s}_{1, k+1}-\hat{s}_{1, k}\right\rangle \nonumber \\
& +h\left\langle{\Delta}_{k-1}, \hat{s}_{2, k+1}-k_2 \hat{m}_{1, k}\right\rangle
\end{flalign}
where the property of subdifferential \eqref{equ:subdiff} has been employed.
Here, we utilize the properties of $\theta$-monotone operators. An operator $\mathbf{M}$ is termed $\theta$-monotone if there exists a function $\theta: \mathbb{R}^n\times \mathbb{R}^n \to \mathbb{R}^n$ such that $\theta(x,y)=\theta(y,x)$, and for any two pairs $(v_i,w_i)\in \mathrm{Gph}\mathbf{M}$, $i\in \{1,2\}$,
\begin{flalign}\label{equ:theta_monotone}
\theta(v_1,v_2) \|v_1-v_2\|\leq \left\langle v_1-v_2,w_1-w_2 \right\rangle.
\end{flalign}
 where $\mathrm{Gph}\mathbf{M}$ denotes the graph of $\mathbf{M}$.

Because $\partial \Psi_1(\cdot)$ is a $\theta$-strongly monotone operator with $\theta(x,y)=\beta \|x-y\|$, $\beta>0$, \eqref{equ:Lyap02} follows that
\begin{flalign}\label{equ:Lyap03}
 \Delta V_{2,k} &\leq-\frac{1}{2}\left\|\hat{s}_{2, k+1}-k_2 \hat{m}_{1, k}-\left(\hat{s}_{2, k}-k_2 \hat{m}_{1, k-1}\right)\right\|^2 \nonumber \\
&-  \frac{k_2}{h}\theta(\hat{s}_{1, k+1},\hat{s}_{1, k}) \|\hat{s}_{1, k+1}-\hat{s}_{1, k}\| \nonumber \\
& +h\|\Delta_{k-1}\| \| \hat{s}_{2, k+1}-k_2 \hat{m}_{1, k}\|  \nonumber \\
&= -\frac{1}{2}\left\|\hat{s}_{2, k+1}-k_2 \hat{m}_{1, k}-\left(\hat{s}_{2, k}-k_2 \hat{m}_{1, k-1}\right)\right\|^2  \nonumber \\
&-(k_2 \theta(\hat{s}_{1, k+1},\hat{s}_{1, k})-h\delta_3)\| \hat{s}_{2, k+1}-k_2 \hat{m}_{1, k}\|,
\end{flalign}
from which one can conclude that $\Delta V_{2,k}\leq 0$ whenever $\theta(\hat{s}_{1, k+1},\hat{s}_{1, k})\geq h\delta_3/k_2 $. Therefore, there exits a stepping number $k^\ast\in \mathbb{N}$ and $\epsilon_1(k_2)>0$ such that 
\begin{flalign}\label{equ:lyapunv04}
\|\hat{s}_{2, k+1}-k_2 \hat{m}_{1, k}\|\leq h\epsilon_1(k_2),\quad \forall k \geq k^\ast
\end{flalign}
From \eqref{equ:hats2a} and \eqref{equ:lyapunv04}, one has $\|\hat{s}_{1,k+1}-\hat{s}_{1,k}\|\leq h^2\epsilon(k_2)$. Furthermore, as $\partial \Psi_1(\cdot)$ is continuously differentiable, as $\|\hat{s}_{1,k+1}-\hat{s}_{1,k}\|\leq h^2\epsilon(k_2)$, one has 
\begin{flalign}\label{equ:lyapunv05}
\|\hat{m}_{1, k}-\hat{m}_{1, k-1}\|\leq h\epsilon_2(k_2)
\end{flalign}
for all $k\geq k^\ast$. Therefore, from \eqref{equ:lyapunv04} and \eqref{equ:lyapunv05}, $\forall k \geq k^\ast$, one has 
\begin{flalign}\label{equ:lyapunv06}
&\|\hat{s}_{2, k+1}-k_2 \hat{m}_{1, k}-(\hat{s}_{2, k}-k_2 \hat{m}_{1, k-1})\|\leq 2h\epsilon_1(k_2), \nonumber  \\
&\Longleftrightarrow \|\hat{s}_{2, k+1}-\hat{s}_{2, k}\|\leq 2h\epsilon_1(k_2)+h\epsilon_2(k_2).
\end{flalign}
Therefore, from \eqref{equ:hats2}, one has 
\begin{flalign}\label{equ:lyapunv07}
\| \hat{m}_{2,k}\|\leq (2h\epsilon_1(k_2)+h\epsilon_2(k_2)+\delta_3)/k_3=r,
\end{flalign}
from which and the condition $\theta(\hat{s}_{1, k+1},\hat{s}_{1, k})\geq h\delta_3/k_2 $, as $k_2$ and $k_3$ increase and are sufficiently large, 
$\| \hat{m}_{2,k}\|=r$ can be achieved for $\forall k\geq k^\ast$ and some $r>0$ such that $r\mathcal{B}\subset \partial \Psi_2(0)$. Due to $\hat{m}_{2,k}\in \partial \Psi_2(\hat{s}_{1, k+1}) $, then one has $\hat{s}_{1, k+1}=0$ for all $k\geq k^\ast+1$. Additionally, because $\partial \Psi_1$ is continuously differentiable, one has $\partial \Psi_1(\hat{s}_{1, k+1})=\hat{m}_{1,k}=0 $ and $\hat{s}_{2, k+1}$ \eqref{equ:hats2} for all $k\geq k^\ast+1$. Therefore, from \eqref{equ:proposed_admin_004} and \eqref{equ:hats2}, one conclude that $\|s_{1,k}\|\leq h^2\delta_3$ and $\|s_{1,k}\|\leq h \delta_3$ for all $k\geq k^\ast+1$.

To reduce the computation complexity of \eqref{equ:proposed_admin_002}, which has to obtain $\hat{m}_{1,k}$ and $\hat{m}_{2,k}$, as done in our previous work\cite{Xiong_2022_TCASII_STA_Semi}, let us
approximate the term $\hat{ m}_{1,k}$ as $ \|s_{1,k}\|^{\frac{1}{2}} \hat{m}_{2,k}$. Then, $u_{s,k}$ in \eqref{equ:proposed_admin_003} can be approximated as follows:
\begin{subequations}\label{equ:proposed_admin_004app}
\begin{flalign}
& u_{s,k}=k_2 \|s_{1,k}\|^{\frac{1}{2}} \hat{ m}_{2,k}+v_{k+1},  \label{equ:proposed_admin_004appa}\\
& v_{k+1}=v_k+hk_3 \hat{ m}_{2,k}, \hat{ m}_{2,k}\in \partial \Psi_2(\hat{s}_{1,k+1}), \label{equ:proposed_admin_004appb}\\
& \hat{s}_{1,k+1}=s_{1,k}-hk_2\|s_{1,k}\|^{\frac{1}{2}}\hat{ m}_{2,k}-h^2k_3 \hat{ m}_{2,k} \label{equ:proposed_admin_004appc}
\end{flalign}
\end{subequations}
which simplifies the computation complexity due to the fact that only one algebraic function with $\Psi_2(\hat{s}_{1,k+1})$ of $\hat{s}_{k+1}$ is to be solved. 

Please note that \eqref{equ:proposed_admin_002}, the implicit Euler discretization of the continuous-time expression \eqref{equ:error_dy03}, is presented in the standard MSTA formulation to demonstrate closed-loop stability for simplicity. While \eqref{equ:proposed_admin_004} approximate \eqref{equ:proposed_admin_002} to reduce the complexity for real-time computation.
In practices, to obtain the explicit expression of $\hat{m}_{2,k}$ with faster computation, let us first discretize \eqref{equ:error_dy02} with the implicit Euler method, and then follow the approximation method in \eqref{equ:proposed_admin_004}:
\begin{subequations}\label{equ:proposed_admin_005}
\begin{flalign}
&A_k{s}_{k+1}-M_ks_{k}=-h(f_{e,k}+f_{c,k}+M_ku_{s,k})& \\
& u_{s,k}=k_2 \|s_{k}\|^{\frac{1}{2}} \hat{ m}_{2,k}+v_{k+1},  \label{equ:proposed_admin_005c}\\
& v_{k+1}=v_k+hk_3 \hat{ m}_{2,k}, \hat{ m}_{2,k}\in \partial \Psi_2(\hat{s}_{k+1}), \label{equ:proposed_admin_005d}\\
& A_k\hat{s}_{k+1}\!-\!M_ks_{k}\!=\!-hM_k(k_2 \|s_{k}\|^{\frac{1}{2}} \hat{ m}_{2,k}+\!hk_3 \hat{ m}_{2,k}) & \label{equ:proposed_admin_005a}
\end{flalign}
\end{subequations}
where $A_k=M_k+hC_k+hk_1$ and $s_{1,k}=s_{k}$. Please note \eqref{equ:proposed_admin_005d} follows \eqref{equ:proposed_admin_004appc} to calculate $\hat{s}_{k+1}$. 
\begin{subequations}\label{equ:proposed_admin_0052}
\begin{flalign}
& \hat{m}_{2,k}=-\frac{1}{h\gamma_k}(M_k^{-1}A_k \hat{s}_{k+1}-s_k),  \label{equ:proposed_admin_0052a}\\
& \hat{s}_{k+1}=\mathrm{Prox}_{\mu h \gamma_k\Psi_2}((I_n-\mu M_k^{-1}A_k)\hat{s}_{k+1}+\mu s_k  ) \label{equ:proposed_admin_0052b}
\end{flalign}
\end{subequations}
where $\gamma_k = k_2 \|s_k\|^{1/2} + h k_3$, and $0<\mu <1$ is chosen such that $A_k + A_k^T - \mu A_k^T A_k > 0$. The term $\mathrm{Prox}_{\mu h \gamma_k \Psi_2}(z)$, for all $z \in \mathbb{R}^n$, denotes the proximal map of index $\mu > 0$ for the function $h \gamma_k \Psi_2$ at point $z$ (refer to the definition of the proximal map in \cite[Lemma 2]{Brogliato_2017_TAC}). The algebraic function \eqref{equ:proposed_admin_0052b} can be quickly calculated with the Newton's method.

To avoid the expensive computation of matrix inversion of $M_k^{-1}$, let us again set $k_1=-C_k+\gamma_1M_k$. 
Then, by applying \cite[Lemma 4]{Brogliato_2017_TAC} to \eqref{equ:proposed_admin_005}, one has
\begin{subequations}\label{equ:proposed_admin_0053}
\begin{flalign}
& \hat{m}_{2,k}=-\frac{1}{h\gamma_k}(\beta \hat{s}_{k+1}-s_k),  \label{equ:proposed_admin_0053a}\\
& \hat{s}_{k+1}=\mathrm{Prox}_{\mu h \gamma_k\Psi_2}((1-\mu \beta)\hat{s}_{k+1}+\mu s_k  ) \label{equ:proposed_admin_0053b}
\end{flalign}
\end{subequations}
where $\beta=h\gamma_1+1$, and $0<\mu <1$. 

\subsubsection{Decoupled Manner}
As one special case of \eqref{equ:proposed_admin_005}, i.e., $n=1$, $A_k\in \mathbb{R}$, then one has:
\begin{subequations} \label{equ:sta_bar}
\begin{flalign}
& A_k{s}_{k+1} =M_ks_k-h(f_{e,k}+f_{c,k}+M_ku_{s,k}), \label{equ:sta_bara} \\
& u_{s,k}= {k}_2|\hat{s}_{k+1}|^{\frac{1}{2}}\sgn( \hat{s}_{k+1})+v_{k+1} \label{equ:sta_barc} \\
 & v_{k+1}\in v_{k}+hk_3\sgn(\hat{s}_{k+1}),\label{equ:sta_bard}
\end{flalign}
\end{subequations}
which retrieves to the implicit Euler discretization of STA and $k_4=0$ is typically set. Please note in\cite{Xiong_2022_TCASII_STA_Semi}, $u_{s,k}= {k}_2|{s}_{k}|^{\frac{1}{2}}\sgn( \hat{s}_{k+1})+v_{k+1}$ is called semi-implicit Euler discretization, while in\cite{andritsch2023modified}, $u_{s,k}$ is written as \eqref{equ:sta_barc}.
 From \eqref{equ:proposed_admin_005a}, as done in \eqref{equ:proposed_admin_0053}, for simplicity, let select the gain $k_1=-C_k+\gamma_1M_k$, and the nominal model about $\hat{s}_{k+1}$ is 
\begin{flalign}\label{equ:zhat}
\frac{s_k}{\beta} \in \hat{s}_{k+1}+\frac{h}{\beta}({k}_2|\hat{s}_{k+1}|^{\frac{1}{2}}+hk_3)\sgn( \hat{s}_{k+1})
\end{flalign}
where $\beta=h\gamma_1+1$. This step is important to obtain the second-order accuracy for the STA discretized with implicit Euler methods, as mentioned in \cite{Xiong_2022_TCASII_STA_Semi}. Then, one has $|s_k|=\beta|\hat{s}_{k+1}|+hk_2|\hat{s}_{k+1}|^{1/2}+h^2k_3$ and 
\begin{flalign}\label{equ:z3}
|\hat{s}_{k+1}|^{\frac{1}{2}}=\frac{-hk_2}{2\beta}+\frac{1}{2\beta}\sqrt{h^2k_2^2+4(|\sigma_k|-h^2k_3)}.
\end{flalign}

Then, similar to \cite{Xiong_2022_TCASII_STA_Semi}, to enhance the robustness to disturbances when $\hat{s}_{k+1}=0$, the gain ${k}_2|\hat{s}_{k+1}|^{\frac{1}{2}}$ can be enhanced as ${k}_2|\hat{s}_{k+1}|^{\frac{1}{2}}+h^2k_3$. Then, the implicit Euler discretization of $u_{s,k}$ in \eqref{equ:sta_bar} is as follows:
\begin{subequations}\label{equ:proposed_admin_007}
\begin{flalign}
&u_{s,k}\!=k_2\Phi_1+\!v_{k+1},  v_{k+1}=v_k+hk_3\Phi_2\\
&\Phi_1\in ({k}_2|\hat{s}_{k+1}|^{\frac{1}{2}}+h^2k_3)\mathrm{sgn}(\hat{s}_{k+1}),\Phi_2\in \mathrm{sgn}(\hat{s}_{k+1}).
\end{flalign}
\end{subequations}
From \eqref{equ:zhat}, when $\hat{s}_{k+1}\neq 0$, one has $\mathrm{sgn}(\hat{s}_{k+1})=\mathrm{sign(s_k)}$. When $ \hat{s}_{k+1}=0$, i.e., during the discrete-time sliding surface, from \eqref{equ:zhat}, one has $\hat{s}_{k+1}=0\in s_k-h^2k_3\mathrm{sgn}(\hat{s}_{k+1})$. Then, by using \cite[Eq.(6)]{Kikuuwe_2010_PSMC}, one has, $\hat{s}_{k+1}=s_k-h^2k_3\mathrm{sat}({s_k}/({h^2k_3}))$ where the saturation function $\mathrm{sat}(\cdot)$ is defined as: $\forall z\in \mathbb{R}$, $\mathrm{sat}(z)=z$ if $|z|\leq 1$ and 
$\mathrm{sat}(z)=\mathrm{sign}(z) $ if $|z|>1$. By replacing $\mathrm{sgn}(\hat{s}_{k+1})$ with $\mathrm{sat}({s_k}/({h^2k_3}))$, $|\hat{s}_{k+1}|^{\frac{1}{2}}$ with \eqref{equ:z3} and using $\max(\cdot,\cdot)$ to avoid minus values in the square root function, one can further rewrite \eqref{equ:proposed_admin_007} as follows:
\begin{flalign}\label{equ:sta_last}
\begin{aligned}
  u_{s,k}=&{k}_2\Phi_{1}+v_{k+1},\, v_{k+1} =v_{k}+hk_3\Phi_{2},
\end{aligned}
\end{flalign}
and
\begin{subequations}\label{equ:Phi}
\begin{flalign}\
\Phi_{1}= & \operatorname{sign}\left(s_k\right)\left(\frac{h k_3}{{k}_2} \operatorname{sat}\left(\frac{\left|s_k\right|}{h^2 k_3}\right)-\frac{h {k}_2}{2\beta}+\right. \nonumber \\
& \left.+\frac{1}{2\beta}\sqrt{{h^2 {k}_2^2}+4\max \left(0,\left|s_k\right|-h^2 k_3\right)}\right), \\
\Phi_{2}= & \operatorname{sat}\left(\frac{s_k}{h^2 k_3}\right),
\end{flalign}
\end{subequations}
which is exactly the results of the implicit Euler discretization of STA in \cite[Eq.(11)]{andritsch2023modified}, except for the additional gain $\beta$. Please note that although similar results of \eqref{equ:sta_last}-\eqref{equ:Phi} were shown in \cite{andritsch2023modified}, here it is the first time to show its derivations 
from its original formulation \eqref{equ:sta_bar} for the implicit Euler discretization of the STA, which have not be covered in any previous works\cite{andritsch2023modified,Brogliato_2019_STA}.

To show the stability of the discrete-time closed-loop system interconnected by \eqref{equ:sta_bara} and \eqref{equ:proposed_admin_007} (or its equivalence \eqref{equ:sta_last}-\eqref{equ:Phi}), as done in \eqref{equ:proposed_admin_003}, let us rewrite them as follows:
\begin{subequations} \label{equ:lypunov3}
\begin{flalign}
&  {s}_{1,k+1}\! =\hat{s}_{1,k+1}+hs_{2,k+1} \label{equ:lypunov3a} \\
& s_{2,k+1}\in s_{2,k}-hk_3\Phi_{2}-h\Delta_k,\label{equ:lypunov3d}\\
&  \hat {s}_{1,k+1}\! =\!s_{1,k}\!-hk_2\hat{\Phi}_1-h^2k_3\Phi_{2}
\end{flalign}
\end{subequations}
where $\hat \Phi_1=\partial \Psi_3(\hat{s}_{1,k+1}) $, $\Phi_2=\partial \Psi_4(\hat{s}_{1,k+1}) $, $\Psi_3(\cdot)=\frac{2}{3}|\cdot|^{3/2}+\frac{hk_3}{k_2}|\cdot|+\frac{\gamma_1}{k_2}|\cdot|^2$, and $\Psi_4(\cdot)=|\cdot|$. Please note that $\beta$ in \eqref{equ:zhat} is included in $\hat \Phi_1$, ensuring that \eqref{equ:lypunov3} follows the standard formulation of \eqref{equ:proposed_admin_003}. Consequently, the subsequent proof can be demonstrated similarly from \eqref{equ:Lyap02} to \eqref{equ:lyapunv07}.


\begin{figure}[!t]
	\centering
		\includegraphics[width=0.35\textwidth]{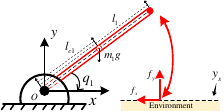}
	\caption{Illustration of one DoF manipulator is accelerated by $f_d$ to impact with unknown environments in terms of stiffness. The parameters are $m_1=5$kg, $l=0.5$m, friction coefficient $\mu=0.1$ and torque saturation $F=3$Nm; }
	\label{fig_manipulator_one-DOF}
\end{figure}



\subsubsection{Implementation of Admittance Controller}
Then, finally, one obtains the final implicit Euler discretization method for \eqref{equ:kikuuwe_adm_proposed}:
\begin{subequations}\label{equ:finally}
\begin{flalign}
&u_{x,k}^\ast \triangleq {(M_x+B_x h)}^{-1}({M_x \dot{q}_{x,k-1}+h\left(f_{c,k}+f_{d,k}\right)}) \label{equ:finallya}\\
& q_{x,k}^\ast \triangleq q_{x,k-1}+h u_{x,k}^\ast, q_{e,k}=q_{x,k}^\ast-q_k \label{equ:finallyb01}  \\
& \dot{q}_{e,k}=[hq_{e,k}\!-\!(q_{x,k-1}\!-\!q_{k-1})]/h, s_k=\dot q_{e,k}+\Lambda q_{e,k} \label{equ:finallyb} \\
& \phi_{a,k} \triangleq \frac{(M_k+{C_k}h)q_k+Bhq_{k-1}}{h^2}+G_k+M_ku_{s,k}  \label{equ:finallyd}\\
&  \phi_{b,k} \triangleq \frac{M_k(q_{x,k-1}+hu_{x,k-1})}{h^2}+\frac{\hat{B}q_{x,k-1}}{h}, \label{equ:finallye} \\
& q_{1,k}^\ast \triangleq q_k+(M_k/h^2+\hat{K})^{-1}(\phi_{b,k}-\phi_{a,k})\label{equ:finallyf} \\
&\tau_k^\ast=(M_k/h^2+\hat{K})(q_{x,k}^\ast-q_{1,k}^\ast), \label{equ:finallyg}\\
&\tau_k=F\operatorname{Proj}(\mathcal{F}; F^{-1}\tau_k^\ast)\label{equ:finallyh} \\
&q_{x,k}=(M_k/h^2+\hat{K})^{-1}\tau_k+q_{1,k}^\ast   \label{equ:finallyi} \\
&  \dot{q}_{x,k}=(q_{x,k}-q_{x,k-1})/h, \label{equ:finallyj}
\end{flalign}
\end{subequations}
where $u_{s,k}$ can be calculated in according with \eqref{equ:proposed_admin_0053} when $u_{s,k}$ is a vector, while it can be calculated with \eqref{equ:sta_last} when it is a scalar. 
Here, $q_{e,k}$ is defined as $q_{x,k}^\ast - q_k$ instead of $q_{x,k} - q_k$, deviating from the definition $q_e = q_x - q$. In \eqref{equ:finallyb}, calculating $s_k$ requires using $q_{x,k}$, but to determine $q_{x,k}$ as shown in \eqref{equ:finallyi}, we need $\tau_k$ and $u_{s,k}$. Therefore, we define $q_{e,k}$ as $q_{x,k}^\ast - q_k$ rather than following the original definition. Note that when $\tau_k^\ast = \tau_k$, it follows from \eqref{equ:finallyg}-\eqref{equ:finallyi} that  $ q_{x,k}^{\ast}= q _ { x , k }$. Thus, replacing $q_{ e , k } =q_{ x, k} ^{\ast }-q_{ k }$ is acceptable for unsaturated cases and helps the admittance controller quickly escape saturated conditions.

In \eqref{equ:kikuuwe_adm_proposed} and its corresponding discrete-time implementation \eqref{equ:finally}, the parameters $f_{c,k}$, $f_{d,k}$, $M_k,C_k, G_k$ represent the values of $f_c$, $f_d$, $M(q)$, $C(q,\dot{q})$ at the time-stepping $k$, which can be obtained from sensors or nominal dynamic models. The parameter $h$ is the sampling period. The SMC controller parameter $F_i$ in $F=\diag(F_1,\cdots, F_n)$ can be determined using the saturation level of the $i$th joint. The parameters $M_x, B_x$ of the proxy system can be designed as $M_x\succ 0$ and $B_xM_x^{-1}$ is Hurwitz. The transient parameter can be accordingly tuned with $\Lambda< 1/h$. The MSTA parameter $k_1=-C_k+\gamma_1M_k$ can be simply designed and then the parameters $\hat{K}=\hat{B}/h+K$, $\hat{B}=B+C_k$, $B=M_k\Lambda+k_1I_n$, $K=(C_k+k_1I_n)\Lambda$ can be identified.subsequently. The parameters $\gamma_1,k_2,k_3$ and $k_4$ are all positive scalars and they can be tuned as the corresponding gains of MSTA by following the guideline of \cite[Algorithm 1]{Moreno_2022_MSTA}.

\begin{figure}[!t]
\centering
  \includegraphics[width=0.40\textwidth,trim=0.6cm 0.8cm 2.2cm 1.9cm,clip]{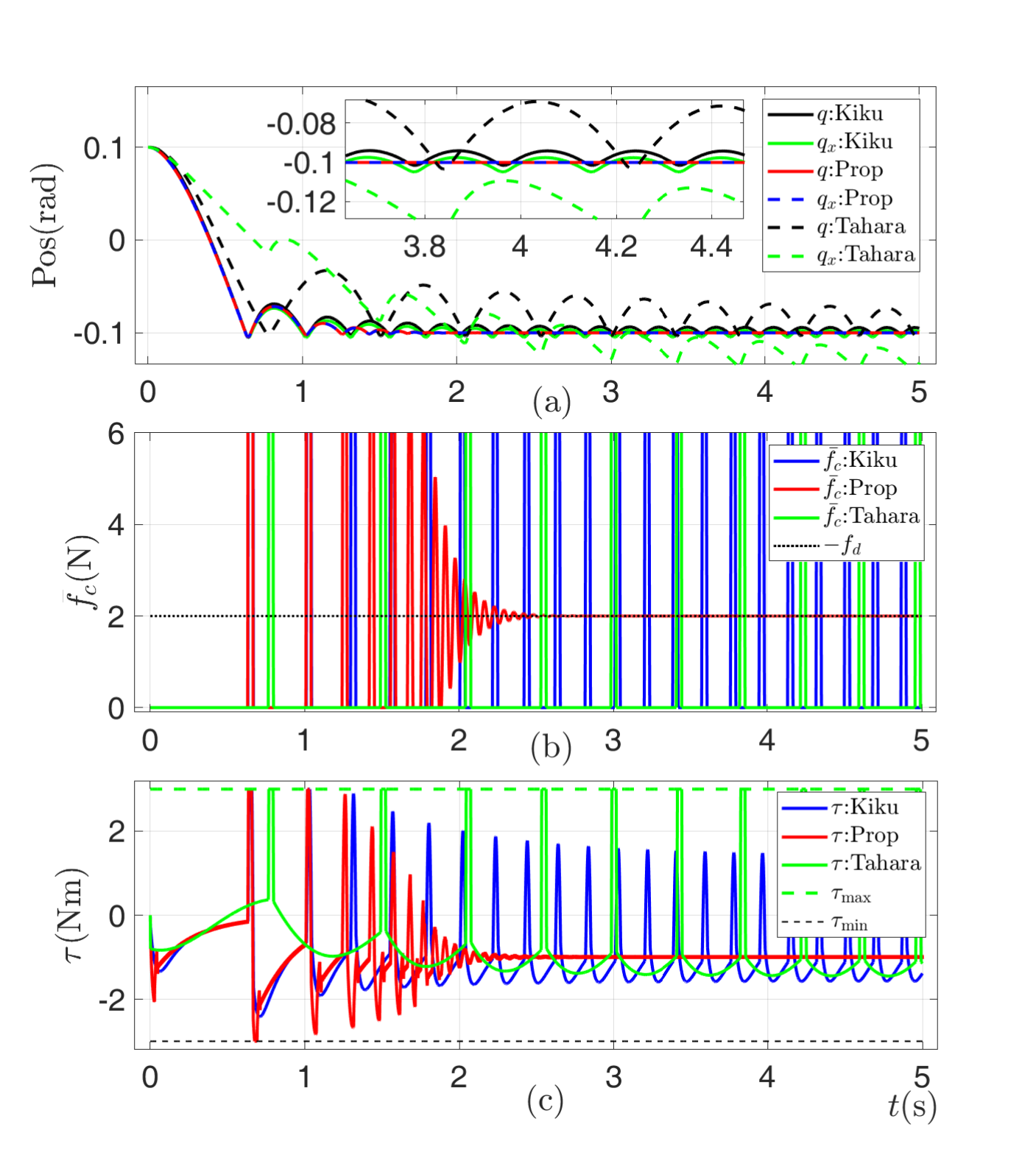}
  \caption{The simulation results of impact-contact scenario in Fig.~\ref{fig_manipulator_one-DOF} with an environment of stiffness $k_s=2\times 10^3$N/m, $f_d=-2$N. The proposed algorithm \eqref{equ:kikuuwe_adm_proposed} denoted as ``Prop", the work in \cite{Tahara_2021_ICRA} denoted as ``Tahara", and the work \cite{Kikuuwe_TRO_2019} denoted by ``Kiku" are implemented and compared with time-stepping size $h=0.001$s; (a) The angular position $q$ of the manipulator and its proxy position $q_x$; (b) The force command $f_d$ and actual contact force $f_c$; (c) The input actuation torque $\tau$ and saturation levels $\tau_{\max}=F$, $\tau_{\min}=-F$; }
  \label{Simulatoin_results}
\end{figure}
\begin{figure}[!t]
	\centering
		\includegraphics[width=0.40\textwidth]{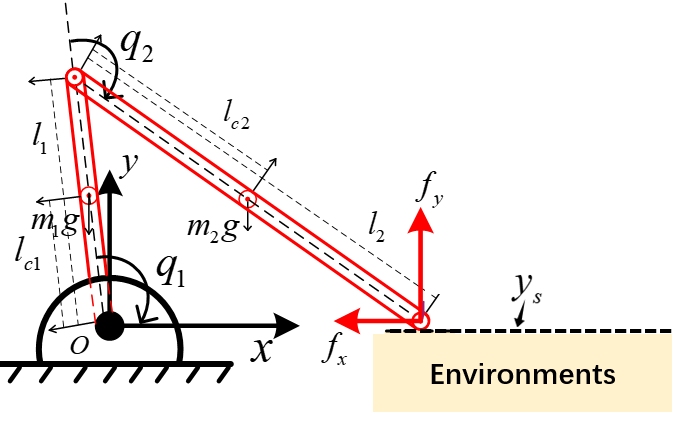}
	\caption{Illustration of two-link planar manipulator is accelerated by $f_d$ to impact with unknown environments in terms of stiffness. The parameters are $m_1=6$kg, $m_2=9$kg,$l_1=0.4$m, $l_2=0.6$m, $J_1=0.32$kg$\cdot$m$^2$, $J_2=1.08$kg$\cdot$m$^2$, friction coefficient $\mu=0.1$ and torque saturations $F_1=3$Nm and $F_2=4$Nm; }
	\label{fig_manipulator_2-DOF}
\end{figure}

\section{Simulations and Experiments}
\label{Sec:impact}

\subsection{Impact-Contact Simulations}
\subsubsection{One-DoF Impact-Contact Force Control}
Fig.~\ref{fig_manipulator_one-DOF} shows the simulation scenario composed of a one-DoF manipulator and a flat surface representing the environment of unknown stiffness, where $l_1 $ is the length of link, $l_{c1} = l_1/2$ is the distance from the origin
to the center of mass of link, and $m_1$ is the mass of centroid of link. The dynamics model of the one-DoF manipulator is given as follows:
\begin{subequations}\label{equ:dynamics_manipulator}
\begin{flalign}
& M\ddot q_1 + C\dot q_1+G = \tau  + f_c,\, G=m_1gl_{c1}\cos(q_1)&\\
& M = {J}_s + {m_1}l_{c1}^2 + 0.2\sin \left( {q_1} \right),  C = 0.1\cos \left( {q_1} \right) &
\end{flalign}
\end{subequations}
where $f_c$ is the contact torque according to the configuration of Fig.~\ref{fig_manipulator_one-DOF}, $M$, $C$, and $G$ represent inertia matrix and Coriolis matrix with uncertainty, and gravity, respectively, ${J}_s\!=\!{{m_1}l_1^2}/3$, and $g$ is the gravity constant.
 The motion of the end-effector along $x$-axis and $y$-axis is caused by the rotation of $q_1$, i.e., $y=l_1\sin(q_1)$ and $x=l_1\cos(q_1)$.

 In \eqref{equ:dynamics_manipulator}, $f_c=J_c^T\bar f_c$ is the contact torque in the joint space with the Jacobian $J_c=[-l_1\sin(q_1), l_1\cos(q_1)]^T$ and the contact force $\bar f_c=[f_{x},f_{y}]^T$ in the Cartesian space. The Coulomb friction along the $x$-axis direction is given by $f_{x}=- \mu {f_{y}}{\rm sign}( {\dot x})$, where $\mu=0.1$ is the friction coefficient, and $f_{c,y}=\max(0,k_s(y_s - y) )$ represents the impact force along $y$-axis direction. Here, $k_s$ denotes stiffness of the interaction environment, and  $y_s=-l\sin(0.1)$m indicates the location of environment.
 When the end-effector comes into contact with surface at location $y_s$ along $y$-direction, it experiences an upward support force $f_{y}$. Using admittance controller \eqref{equ:kikuuwe_adm_proposed} to provide any desired force $f_d<0$, the manipulator accelerates to rotate around joint in the clockwise direction as shown in Fig.~\ref{fig_manipulator_one-DOF}.

To compare the controller \eqref{equ:kikuuwe_adm_proposed} fairly with the state-of-the-art admittance control \cite{Kikuuwe_TRO_2019,Tahara_2021_ICRA}, we set the parameters of proxy system $M_x=0.3$ and $B_x=2$ to be identical for both systems, while $M_x=0.3$ and $B_x=5$ is set to achieve its best performance for \cite{Tahara_2021_ICRA}. We assume that only a rough estimation $\hat{M}=0.1$kg is available while $\hat C=0$, and $\hat G=0$ for the feedforward term $M(q) \ddot q_r+ C(q,\dot{q})\dot{ q}_r+G(q)$, i.e., $\hat{M}\ddot{q}_r$, in \eqref{equ:kikuuwe_adm_proposed}, and $\hat{M}\ddot{q}_x$ in \cite[Eq.(27c)] {Kikuuwe_TRO_2019}. In addition, $u_s$ was calculated with \eqref{equ:sta_last} and we set other parameters as follows:  $k_1=30$,  $k_2 =11.6$, $k_3 =66$, and $k_4=0$ for our proposed algorithm (\eqref {equ:kikuuwe_adm_proposed}), while setting $\Lambda =10,$  $L =0,$ $B = k_{1} + {\hat M}{\Lambda}$,and ${K} =( k_{1} + \hat{C} ){\Lambda}$ in \cite[Eq.(27)] {Kikuuwe_TRO_2019}, to ensure fairness between them.

\begin{figure}[t]
\centering
  \includegraphics[width=0.45\textwidth,trim=0.8cm 3cm 1.5cm 1.7cm,clip]{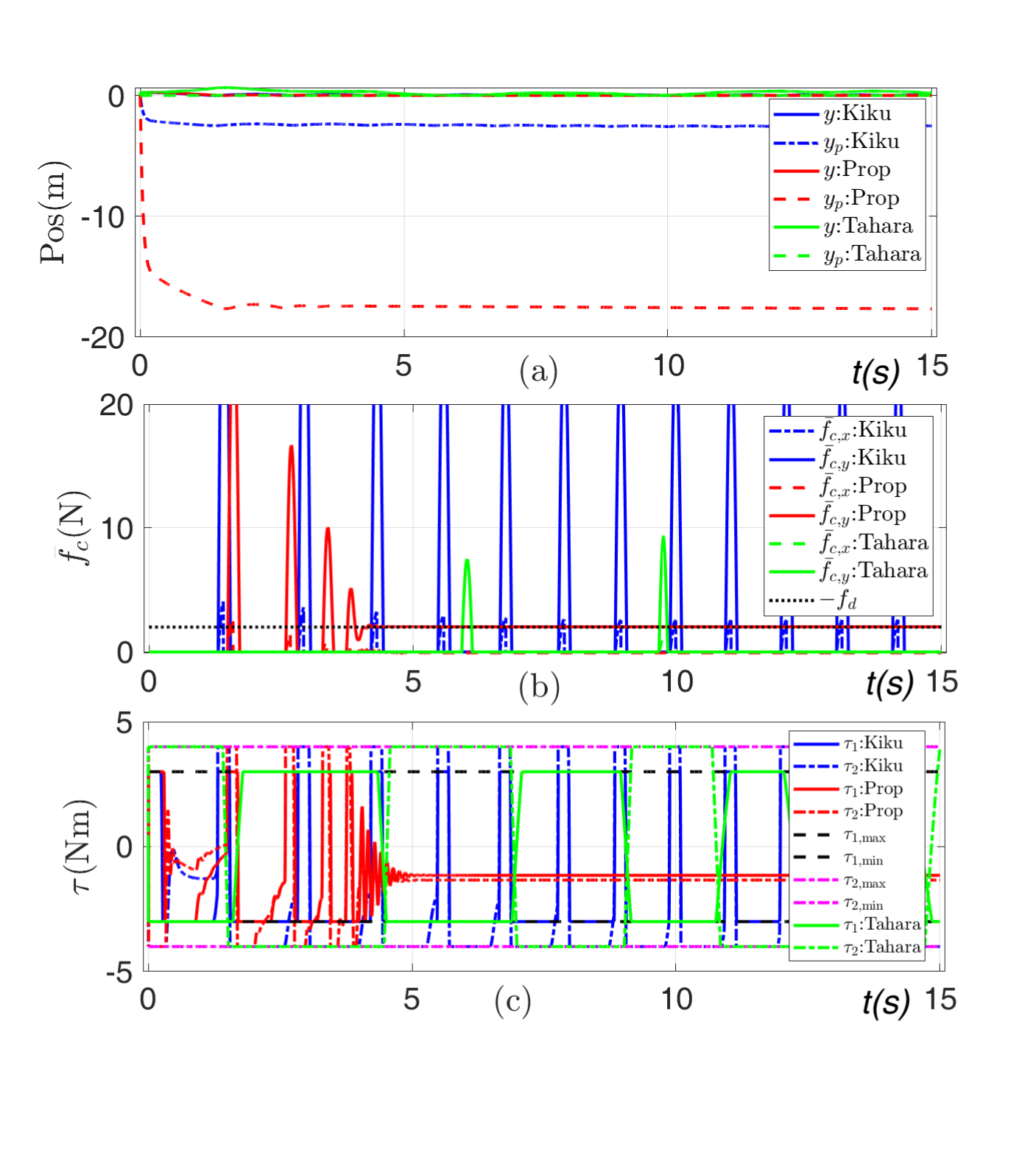}
  \caption{The simulation results of impact-contact scenario in Fig.~\ref{fig_manipulator_2-DOF} with an environment of stiffness $k_s=2\times 10^3$N/m, $f_d=[0,-2]^T$N. The proposed algorithm \eqref{equ:kikuuwe_adm_proposed} denoted as ``Prop", and the work \cite{Kikuuwe_TRO_2019} denoted by ``Kiku" are implemented and compared with time-stepping size $h=0.001$s; (a) The position $y$ of the end-effector of manipulator and its proxy position $y_p$ along the $y$-axis direction in the Cartesian space; (b) The force command $f_d$ and actual interaction force $\bar{f}_c$ along the $x$-axis (friction) and $y$-axis (contact force), i.e., $\bar{f}_{c,x}$ and $\bar{f}_{c,y}$ in the Cartesian space; (c) The input actuation torque $\tau$ of two joints and saturation levels $\tau_{1,\max}=F_1$, $\tau_{1,\min}=-F_1$, $\tau_{2,\max}=F_2$, $\tau_{2,\min}-F_2$; }
  \label{Two_DOF_Simulatoin_results}
\end{figure}

The results of the impact simulation are presented in Fig.~\ref{Simulatoin_results}. In Fig.~\ref{Simulatoin_results}(a), it can be observed that after impact contact, there is a deviation between the proxy position $q_x$ and robot position $q$ for both the algorithm ``Kiku", and the proposed algorithm. The robustness of our proposed controller \eqref{equ:kikuuwe_adm_proposed} is evident as it achieves the desired force $f_d=-2$N even under rough estimation $\hat M\ddot q_r$ of the model $M(q) \ddot q_r+ C(q,\dot q)\dot q_r+G(q)$ after impacting with high stiff environments.
Fig.~\ref{Simulatoin_results}(b) shows that during post-impact, the state-of-the-art approach ``Kiku" fails to track the desired force leading to oscillations and rebounding of manipulator in terms of position, contact force, and torque. Increasing parameter $K$ in \cite[Eq.(27)]{Kikuuwe_TRO_2019} does not improve this situation; instead, it worsens or maintains oscillation levels during simulations. Our simulations found that as stiffness increases (i.e., higher stiff environments), ``Kiku" fails to track desired force resulting in some magnitude oscillations, and ``Tahara" cannot deals with the actuation saturation and also  fails to track desired force with larger resilience forces.
In contrast, \eqref{equ:kikuuwe_adm_proposed} compensates for large model uncertainties caused by rough estimation $\hat M\dot q_r$ and differentiable contact force $f_c$, contributing significantly to robustness exhibited by our proposed controller \eqref{equ:kikuuwe_adm_proposed}.

\subsubsection{Two-DoF Impact-Contact Force Control}
As shown in Fig.~\ref{fig_manipulator_2-DOF}, let us consider the impact-contact force control of a two-link planar manipulator, with its dynamic model detailed in \cite[Sec. VII]{Brogliato_2017_TAC}. The saturation levels for the two joints are set as $F=\mathrm{diag}[F_1, F_2]$ with $F_1=3$ Nm and $F_2=4$ Nm. For simplicity, we set $M_x$ and $B_x$ in \eqref{equ:kikuuwe_adm_proposed} to $M_x=\mathrm{diag}[0.5, 0.5]$ and $B_x=\mathrm{diag}[1, 1]$ and the feedforward term $M(q) \ddot q_r+ C(q,\dot{q})\dot{ q}_r+G(q)$ with a rough estimation $\hat{M}=\mathrm{diag}[0.2, 0.2]$, $\hat C=\mathrm{diag}[20, 20]$. The contact force is $\bar f_c=[f_{c,x},f_{c,y}]^T$ in Cartesian space. Fig.~\ref{Two_DOF_Simulatoin_results} shows the force tracking results of after contacting with an environment of high stiffness. Again, after impact-contact and under torque saturation, both ``Tahara" from \cite{Tahara_2021_ICRA} and ``Kiku" from \cite{Kikuuwe_TRO_2019} failed to track the desired contact force $f_d=-2$ N for the two-link manipulator case (see Fig.~\ref{Two_DOF_Simulatoin_results}(b)). The algorithm ``Tahara" cannot handle actuation saturation (see Fig.~\ref{Two_DOF_Simulatoin_results}(c)). In addition, $u_s$ was calculated with \eqref{equ:MSTA_Explicit} for simplicity, and other parameters $k_1$,  $k_2$, $k_3$, $\Lambda $, $B, K$ are set the same as in the case of one DoF case to ensure fairness comparisons. Similar to the case in Fig.~\ref{Simulatoin_results}, Fig.~\ref{Two_DOF_Simulatoin_results} shows the robustness of our proposed controller \eqref{equ:kikuuwe_adm_proposed} as it achieves the desired force $f_d=-2$N along the $y$-axis even under torque saturation and rough estimation of dynamic model $\hat M$ and $\hat{C}$ after impacting with environments.

\subsection{Impact-Contact Experiments}
\label{Sec:experiments}
\subsubsection{Experimental Set-up}
\begin{figure}[b]\centering
  \includegraphics[width=0.8\columnwidth]{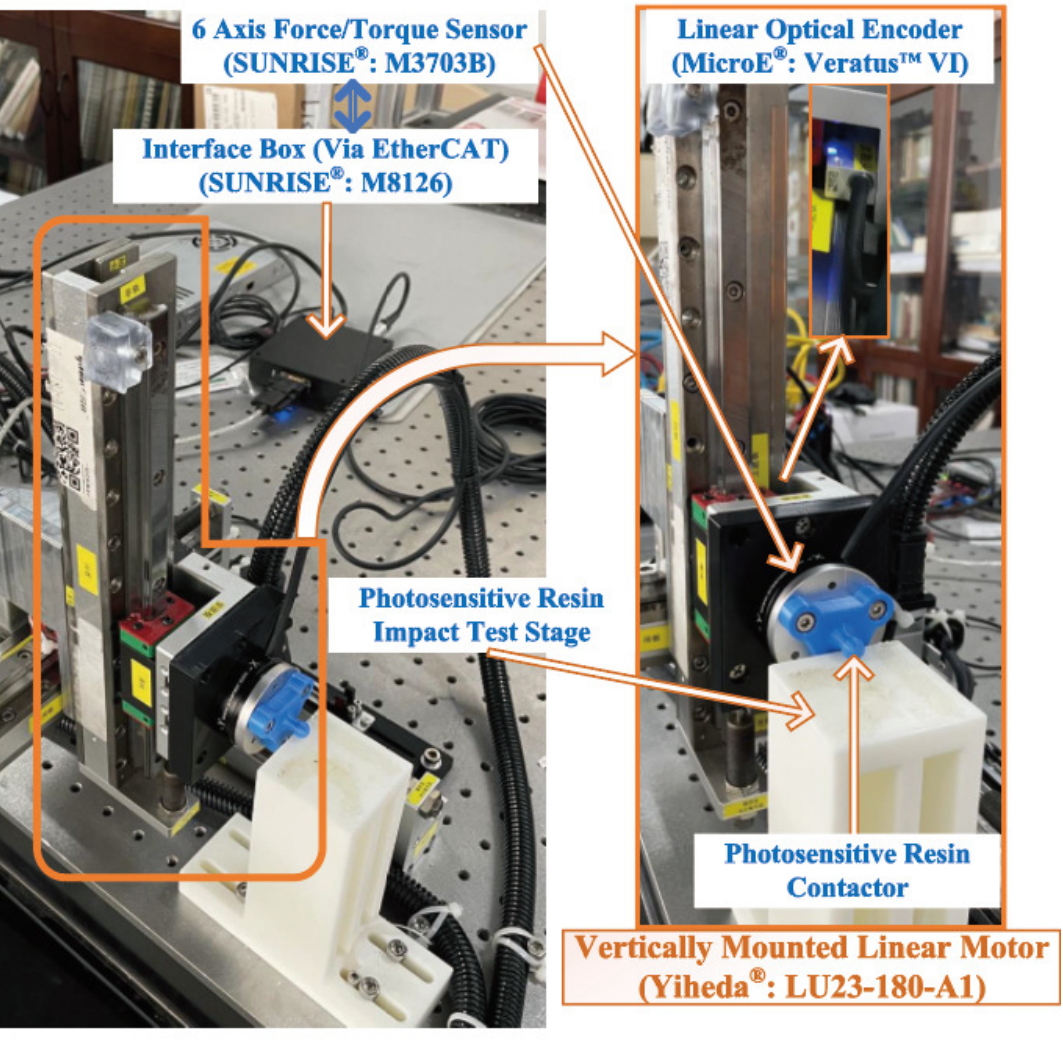}
  \caption{The experimental hardware included a linear motor with an end-effector (a blue photosensitive resin contactor), a force sensor, and an encoder. This end-effector was used to impact the white photosensitive resin impact test stage, which represented various environments.}
  \label{Exp_Platform}
\end{figure}

Fig.~\ref{Exp_Platform} displays the experimental platform and its hardware components of the linear motor system, which is a crucial element in high-speed and high-precision automatic systems like Surface-Mount Technology (SMT) machines and automated assembly machines. The linear motor system comprises several modules such as the linear motor, driver, moving stage, linear bearing rail, linear optical encoder, interface box, 6-axis F/T sensor and Ethernet Control Automation Technology (EtherCAT) controller. Tab.~\ref{model} provides detailed lists of these main components while Fig.~\ref{diff_materials} presents a schematic diagram of the platform shown in Fig.~\ref{Exp_Platform}.

A blue 3D-printed end-effector, depicted in Fig.~\ref{Exp_Platform}, is attached to a force sensor and interacts with varying levels of stiffness in its surroundings. The force sensor is mounted on the slider of a linear bearing rail that can travel up to $180$mm. A linear optical encoder installed on the rail measures the position of the moving stage with a resolution of 0.5$\mu$m, and sends filtered position signals directly back to the linear motor driver for feedback.

 The model of the linear motor system is as follows:
\begin{equation}
M\ddot{q}+C\dot{q}=\kappa u-Mg+F_f+f_c
\label{eq.linear_motor_system_model}
\end{equation}
where $\kappa $ is driver gain, $u$ is the control input, $q$ is the position of the linear motor, $M$ is the mass of the moving stage, $Mg$ denotes the gravitational force, $C$ is the viscous coefficient, and $f_c$ is the contact force measured by the force sensor. The force $F_f$, including the friction force and cogging force, is typical nonlinear and difficult to obtain the exact model, and thus is treated as an external disturbance. The values for parameters $M, C,\kappa$ can be found in our previous work \cite{Xiong_2023_TIE}.


%

\begin{table}[!t]
\small
  \renewcommand{\arraystretch}{1.3}
  \caption{Major Components of the Linear Motor System}
  \centering
  \label{model}
  \resizebox{\columnwidth}{!}{
    \begin{tabular}{l l l}
      \hline\hline \\[-3mm]
      \multicolumn{1}{l}{Components} & \multicolumn{1}{l}{Manufacture} & \multicolumn{1}{l}{\pbox{20cm}{Model}}  \\[1.6ex] \hline
      Six-axis F/T Sensor & Sunrise & M3703B \\
      Interface Box of the F/T Sensor & Sunrise & M8126 \\
      Linear Motor  & Yiheda & LU23-180-A1 \\
      Linear Motor Driver  & Panasonic & MBDLN25BL \\
      Linear Optical Encoder  & MicroE & Veratus$^{\text{TM}}$ VI \\
      EtherCAT Controller  & iManifold & iMF-HG-STD/WD-01

      \\ [1.4ex]
      \hline\hline
    \end{tabular}
   }
\end{table}

\subsubsection{Impact-Contact Experiments}
\label{sec.exp}


\begin{figure}[t]
  \centering
  \centerline{\includegraphics[width=\columnwidth,trim={0cm 0cm 0 0cm},clip]{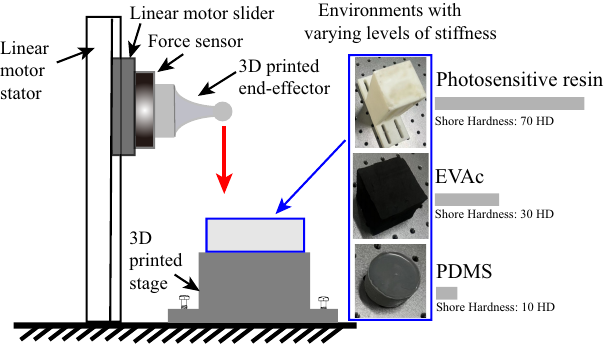}}
  \caption{ The schematic diagram of the platform in Fig.~\ref{Exp_Platform} for experiments of impacting with blocks of three different materials. The stiffness of these materials—photosensitive resin, EVAc, and PDMS—is indicated by their shore hardness values: approximately 70HD, 30HD, and 10HD respectively. Thus, photosensitive resin is the stiffest, followed by EVAc, and then PDMS. }
  \label{diff_materials}
\end{figure}

\begin{figure*}[!t]
\centering
\subfigure[Experimental results of impacting with the block of photosensitive
 resin (3D printing materials) in Fig.~\ref{diff_materials}. ]{
  \includegraphics[width=0.90\textwidth,trim={0cm 0cm 0cm 0cm},clip]{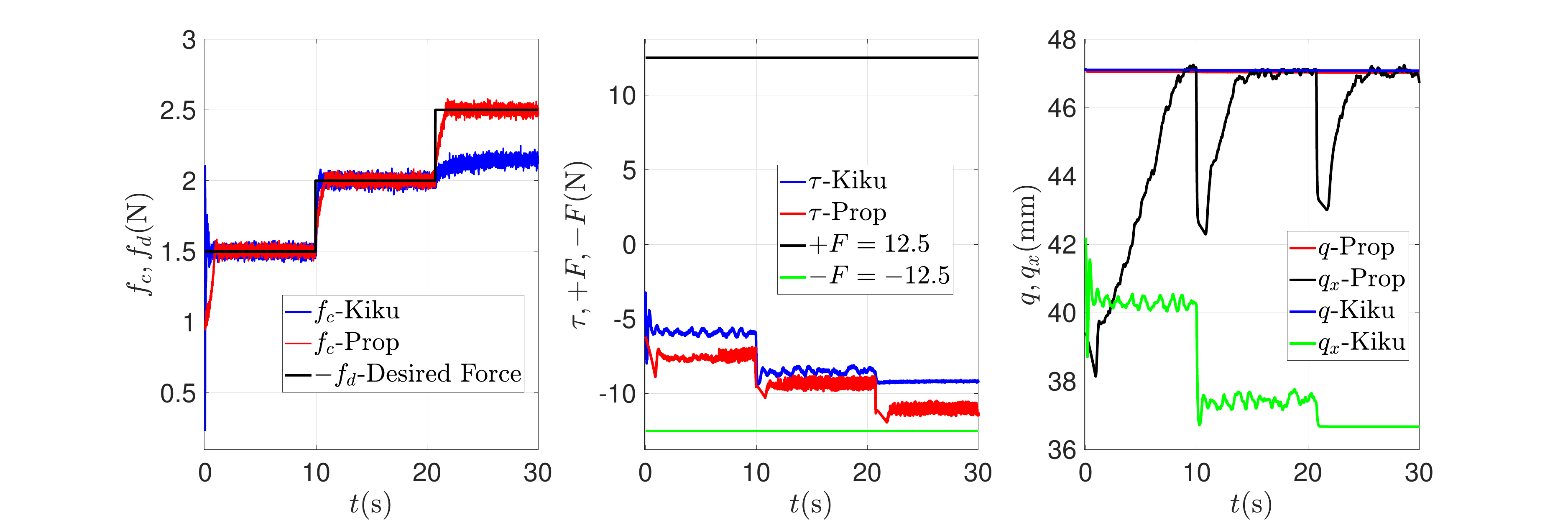}}

\subfigure[Experimental results of impacting with the block of EVAC in Fig.~\ref{diff_materials}.]{
  \includegraphics[width=0.90\textwidth,trim={0cm 0cm 0cm 0cm},clip]{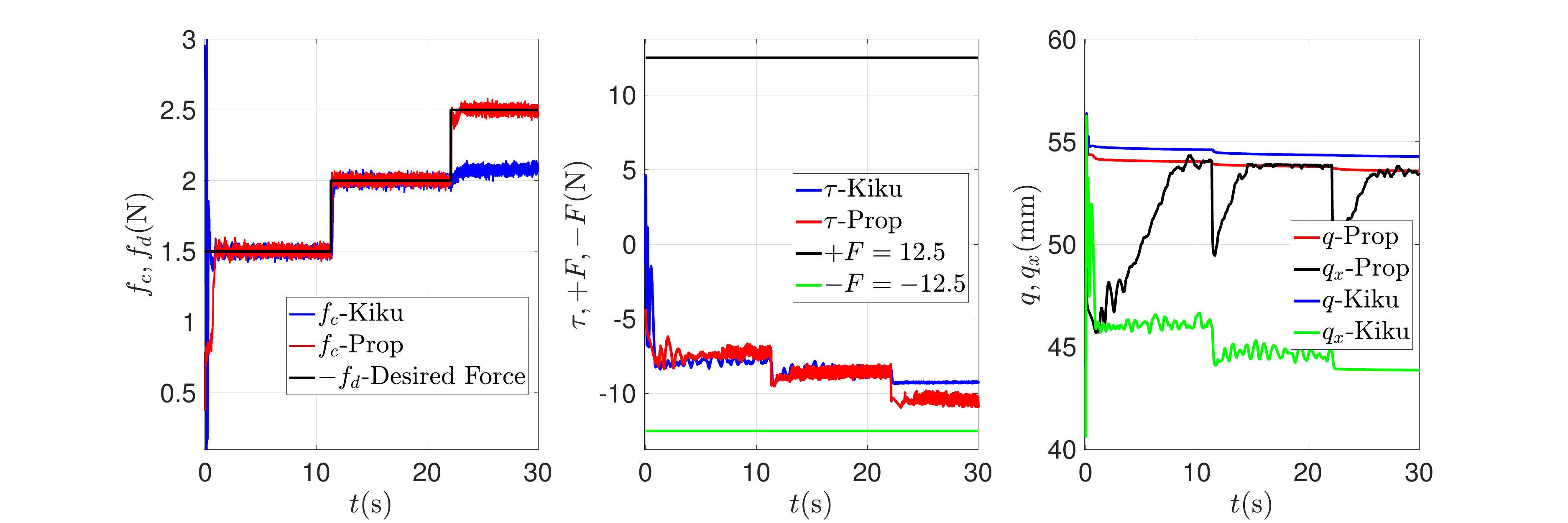}}

  \subfigure[Experimental results of impacting with the block of PDMS in Fig.~\ref{diff_materials}]{
  \includegraphics[width=0.90\textwidth,trim={0cm 0cm 0cm 0cm},clip]{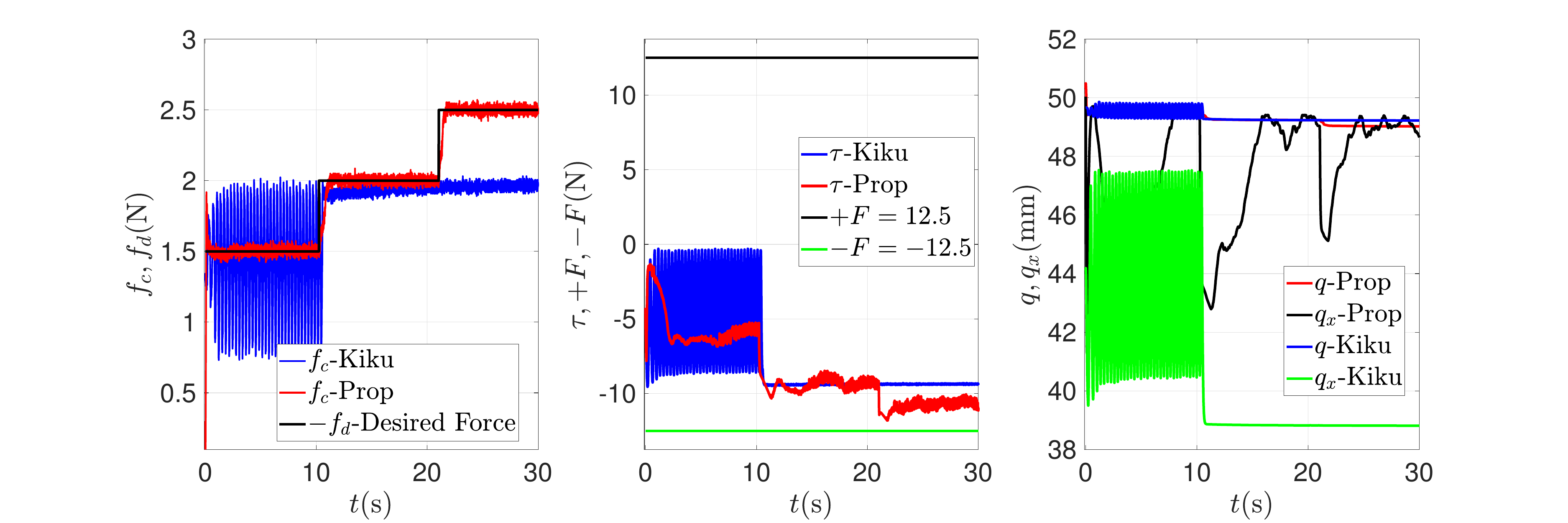}}
\caption{Results of the experiments of impacting to different environments of different levels of stiffness with initial velocity $\dot{q}=0.04$m/s. During these experiments, the parameters were kept constant: $\hat{M}=0.22$Kg, $\hat{C}=\hat{G}=0$, $k_1=60$, $k_2=22.25$, $k_3=242$, $\Lambda=10$ for \eqref{equ:kikuuwe_adm_proposed}, denoted by ``Prop", and $\hat{M}=0.22$Kg, $K=320, B=60, L=310$ for the internal PID has been fine tuned with the admittance control proposed by Kikuuwe \cite{Kikuuwe_TRO_2019}, denoted by ``Kiku", for impacting to the block of photosensitive resin. For both controllers, the values of parameters of $h=4$ms, $M_x=0.2$Kg, $B_x=4$m/s$^2$, and $F=12.5$N were kept the same for principle of fair comparisons.}\label{fig.Exp_3D-01}
\end{figure*}

In Fig.~\ref{diff_materials}, three blocks of different materials with obviously different levels of stiffness were used to represent the contact environments, i.e., the photosensitive Resin of 3D-printing material, the Ethylene-Vinyl Acetate (EVAc), and the Polydimethylsiloxane (PDMS), of which the stiffness is reduced gradually. To avoiding any damage to the hardware, during each trial, the end-effector in Fig.~\ref{diff_materials} was controlled to move with a constant velocity $\dot{q}=0.04$m/s in the vertical direction to impact to each block on the top of the 3D printed stage. After contact and $f_c\neq 0$, the linear motor was switched from the velocity control to the force control. For both the state-of-the-art  \cite{Kikuuwe_TRO_2019} and the proposed one \eqref{equ:kikuuwe_adm_proposed}, their parameters were first tuned sepearately based on the experiments of the photosensitive resin of 3D-printing material to achieve their respective best performances for $f_d=-1.5$N, and then fixed for the other materials and other force commands $f_d=-2$N and $f_d=-2.5$N to show their robustness to changed environments of stiffness and changed force commands.

\par


%
%
%
%
%
Fig.~\ref{fig.Exp_3D-01} shows the contact force tracking results with the proposed algorithm \eqref{equ:kikuuwe_adm_proposed}, denoted as ``Prop", and the admittance control proposed by R. Kikuuwe \cite{Kikuuwe_TRO_2019}, denoted as ``Kiku", after impacting to the environments of three different stiffness levels, 3D printing material, EVAc, and PDMS, respectively, with the same velocity $\dot{q}=0.04$m/s.
As shown in Fig.~\ref{fig.Exp_3D-01}(a), after a transient response of large impact force due to the impact, the ``Kiku" successively tracks the desired contact force $-f_d=1.5$N. It also tracks the desired force $-f_d=2.0$N with the fixed parameters that were tuned for the case of $-f_d=1.5$N. However, when the desired force changes from $-f_d=2.0$N to $-f_d=2.5$N, the ``Kiku" fails to track the force command. However, the ``Prop" successfully the desired contact force $-f_d=1.5$N, $-f_d=2.0$N, and $-f_d=2.5$N, respectively, with the fixed parameters that were tuned for the case of $-f_d=1.5$N.
The obvious difference is observed in the results of $q_x$ and $q$, where $q_x$ has a larger deviation from $q$ for the ``Kiku" method.

\par
In Fig.~\ref{fig.Exp_3D-01}(b), the end-effector was impacted to the EVAC material with the same vertical velocity. It shows that the ``Kiku" algorithm exhibited similar behavior to that shown in Fig.~\ref{fig.Exp_3D-01}(a). The actuation force $\tau$ of the ``Kiku" algorithm successfully tracked desired forces $-f_d=1.5$N and $-f_d=2$N, respectively, indicating some level of robustness to changes in environment and force commands. However, for a desired force of $-f_d=2.5$N, the ``Kiku" algorithm failed to track it. In contrast, as shown in Fig.~\ref{fig.Exp_3D-01}(b), the end-effector using the ``Prop" algorithm successfully tracked desired forces $-f_d=1.5$N, $-f_d=2$N, and $-f_d=2.5$N even when there were changes in environment conditions; this demonstrates high robustness of the proposed algorithm.
\par

Fig.~\ref{fig.Exp_3D-01}(c) shows that after the impact to the PDMS material with the same vertical velocity, the end-effector with ``Kiku" is oscillated with a large magnitude and the contact force cannot converge to the desired force $-f_d=1.5$N, similar to the simulation results in Fig.~\ref{Simulatoin_results}. The actuation force $\tau$ is bounced even though the actuation torque is not saturated. In contrast, after a transient response, the end-effector with ``Prop" quickly keeps a persistent contact with the PDMS material and the contact force is equivalent to the desired force $-f_d=1.5$N, and the actuation force is within the saturation levels $\pm F$. Fig.~\ref{fig.Exp_3D-01}(c) shows that, due to the set-valued STA, the proxy position $q_x$ closely follows the actual position $q$ of the end-effector, while $q_x$ and $q$ of ``Kiku" show significant oscillations, which further results in the oscillations in the contact force $f_c$ and actuation force $\tau$.
It shows that, with fixed parameters in Fig.~\ref{fig.Exp_3D-01}, the ``Kiku" cannot adapt to the changed environment of different stiffness and changed force commands, while the proposed algorithm is more robust to their changes and uncertainties.

%


\section{Conclusion}
\label{sec:conclusion}
%

This paper proposes a new admittance control strategy that utilizes two set-valued feedback loops. The outer loop uses set-valued first-order SMC to address unsafe behaviors after actuation saturation, while the inner loop leverages the differentiability of impact-contact force make the admittance control robust to changes in unknown stiffness environments. Additionally, this paper presents a discrete-time realization of the new structure of admittance control based on implicit Euler discretization. Results show that compared with state-of-the-art control strategies in literature, our proposed strategy is more robust and adaptable to varying stiffness environments.

Future studies should focus on extending this approach to combine learning-based methods to learn the dynamics and environment features. To improve force control accuracy and adaptability during interactions with unknown environments, online tuning for estimating plant parameters (e.g., dynamic models) can be done using well-developed deep neural networks (DNNs) or reinforcement learning (RL). Furthermore, extending the proposed algorithm to multiple DoF cases with multi-variable SMC such as multip-variable STAs is crucial for industrial and automation tasks requiring operations in Cartesian space with different torque limits for multiple joints.


%

%
%
%
%
%
%

\ifCLASSOPTIONcaptionsoff
  \newpage
\fi

\bibliographystyle{IEEEtran}
\bibliography{cas-refs}

\end{document}